\documentclass[a4paper,11pt]{article}
\pdfoutput=1
\usepackage{jcappub}
\usepackage{amssymb}
\usepackage{graphicx}
\usepackage{amsmath}
\usepackage{hyperref}
\usepackage{tensor}
\usepackage[utf8]{inputenc}
\usepackage[normalem]{ulem}

\newcommand{\dif}{\mathrm{d}}
\newcommand{\me}{\mathrm{e}}
\newcommand{\mi}{\mathrm{i}}
\newcommand{\bm}{\mathbf}

\title{\boldmath{Resonant multiple peaks in the induced gravitational waves}}

\author[a]{Rong-Gen Cai,}
\author[b]{Shi Pi,}
\author[c]{Shao-Jiang Wang,}
\author[a,d]{Xing-Yu Yang}

\affiliation[a]{CAS Key Laboratory of Theoretical Physics, Institute of Theoretical Physics, Chinese Academy of Sciences, P.O. Box 2735, Beijing 100190, China}
\affiliation[b]{Kavli Institute for the Physics and Mathematics of the Universe (WPI), The University of Tokyo Institutes for Advanced Study, The University of Tokyo, Kashiwa, Chiba 277--8583, Japan}
\affiliation[c]{Tufts Institute of Cosmology, Department of Physics and Astronomy, Tufts University, 574 Boston Avenue, Medford, Massachusetts 02155, USA}
\affiliation[d]{School of Physical Sciences, University of Chinese Academy of Sciences, No.19A Yuquan Road, Beijing 100049, China}

\emailAdd{cairg@itp.ac.cn}
\emailAdd{shi.pi@ipmu.jp}
\emailAdd{schwang@cosmos.phy.tufts.edu}
\emailAdd{yangxingyu@itp.ac.cn}

\abstract{We identify analytically a multiple-peak structure in the energy-density spectrum of induced gravitational waves (GWs) generated at second-order from a primordial scalar perturbations also with multiple ($n$) peaks at small scales $k_{*i}$.
The energy-density spectrum of induced GWs exhibits at most $C_{n+1}^2$ and at least $n$ peaks at wave-vectors $k_{ij}\equiv(k_{*i}+k_{*j})/\sqrt{3}$ due to resonant amplification, and, under the narrow-width approximation, it contains an universal factor that can be interpreted as a result of momentum conservation.
We also extend these discussions to the case of non-Gaussian perturbations.
}

\begin{document}
\maketitle

\section{Introduction}\label{sec:int}

The current march on the detections of gravitational waves (GWs) with astrophysical origin from compact binary merger\cite{Abbott:2016blz,TheLIGOScientific:2017qsa} has renewed rich perspectives of other sources of GWs preferably from the early Universe\cite{Caprini:2018mtu} that could probe fundamental physics at an unprecedented level\cite{Cai:2017cbj}.
For example, the primordial GWs from inflation era\cite{Guzzetti:2016mkm} could fix the benchmark scale of inflation\cite{Bartolo:2016ami} and possibly rule out\cite{Boyle:2003km} other alternative of cosmological scenarios\cite{Gasperini:2002bn,Lehners:2008vx,Battefeld:2014uga,Cai:2014bea,Brandenberger:2016vhg}; the GWs from preheating/reheating era\cite{Allahverdi:2010xz,Amin:2014eta} could further constrain the inflation model; the GWs from cosmological first-order phase transition\cite{Binetruy:2012ze,Caprini:2015zlo} could pave the road beyond the standard model (SM) of particle physics; and the GWs from topological defects\cite{Binetruy:2012ze} such as cosmic string could be the first smoking gun for string theory.

Recently, the renewed interest\cite{Ananda:2006af,Baumann:2007zm,Saito:2008jc,Saito:2009jt,Garcia-Bellido:2017aan,Ando:2017veq,Espinosa:2018eve,Kohri:2018awv,Cai:2018dig,Bartolo:2018evs,Bartolo:2018rku,Unal:2018yaa,Byrnes:2018txb,Inomata:2018epa} in the induced GWs from primordial scalar perturbations has drawn a lot of attention.
Although the primordial scalar perturbations on large scales have been elaborately probed in a well-established manner\cite{Akrami:2018odb}, the primordial tensor perturbations are still at large from the current scope of detections.
Since the primordial tensor perturbations on CMB scales are small due to the current constraint for the tensor-to-scalar ratio~\cite{Ade:2018gkx,Matsumura:2013aja,Li:2017drr}, it is difficult to be detected in the near future\cite{Crowder:2005nr,Corbin:2005ny,Kawamura:2006up,Kawamura:2011zz}.
However, even though the scalar and tensor perturbations are decoupled at first order in perturbation theory, they are coupled at second order and induced GWs could be sourced by two scalar perturbations in the radiation dominated (RD) universe\cite{Ananda:2006af,Baumann:2007zm}.
Furthermore, if the scalar perturbations are peaked at some small scales which will not affect the well-constrained density perturbations on CMB scales, the induced GWs could be large and detectable in pulsar timing array or future interferometers\cite{AmaroSeoane:2012km,AmaroSeoane:2012je,Audley:2017drz,Guo:2018npi,Luo:2015ght,Crowder:2005nr,Corbin:2005ny,Kawamura:2006up,Kawamura:2011zz,Hobbs:2009yy,Carilli:2004nx}.
Such enhanced scalar perturbations at small scales could also lead to the formation of primordial black holes (PBHs) at the horizon reentry of the corresponding wavelengths\cite{Hawking:1971ei,Carr:1974nx,Carr:1975qj}, which could serve as an appealing candidate for dark matter (DM)\cite{Carr:2009jm,Carr:2016drx,Carr:2017jsz} as well as explaining the large merger rate of binary black holes observed in LIGO detections of GWs\cite{Kashlinsky:2016sdv,Bird:2016dcv,Clesse:2016vqa,Sasaki:2016jop}.

There are many works on the induced GWs recently\cite{Garcia-Bellido:2017aan,Ando:2017veq,Espinosa:2018eve,Kohri:2018awv,Cai:2018dig,Unal:2018yaa,Inomata:2018epa}.
\cite{Garcia-Bellido:2017aan,Ando:2017veq} investigate the possiblity of detecting the induced GWs from those inflationary models that could also generate PBHs as DM.
\cite{Espinosa:2018eve} explores the possibility of detecting the induced GWs from SM due to Higgs meta-stability during inflation.
\cite{Kohri:2018awv} makes a great progress in analytically solving the equation of motion for the induced tensor perturbation.
In\cite{Cai:2018dig} some of the authors of this paper extrapolate the scalar perturbation to be non-Gaussian, and forecast a distinctive observational perspective in the induced GWs for such non-Gaussianity.
They further claim that if PBHs can serve as all the DM in the current affordable window $M_\text{PBH}\sim10^{20}$g to $10^{22}$g, the induced GWs must be detectable by LISA like interfereometers.
\cite{Byrnes:2018txb,Inomata:2018epa} further constrain the curvature perturbations at small scales from the induced GWs probed by the existing and planned GW experiments.
In these works, we noticed that the shape of the energy-density spectrum of induced GWs is sensitive to the shapes and positions of the peaks in the scalar perturbation.
This motivates us to study the GWs induced by multiple peaks, which can be easily generated by inflation model with multiple fields or multiple inflection points\cite{Gao:2018pvq,Cheng:2018yyr}.
We identify a multiple-peak structure of induced GWs from primordial scalar perturbations with multiple-peak.

The outline for this paper is as follows: In section~\ref{sec:GWs}, the formalism of induced GWs is reviewed for the clarity of our notation; In section~\ref{sec:delta}, we obtain the energy-density spectrum of induced GWs from Gaussian scalar perturbations with $\delta$-peak, and a multiple-peak structure is analytically identified; In section~\ref{sec:sigma}, we find the same multiple-peak structure for the energy-density spectrum of induced GWs from Gaussian scalar perturbations with Gaussian peak.
In section~\ref{sec:NGGWs}, we further extend our discussion into the case of non-Gaussian scalar perturbations.
The section~\ref{sec:con} is devoted to conclusion.

\section{Induced GWs from Gaussian scalar perturbations}\label{sec:GWs}

To set the notation, we first review the formalism of induced GWs, and we will follow closely the reference\cite{Kohri:2018awv}.

\subsection{Source term}\label{subsec:source}

To compute the induced GWs, one starts with the following metric
\begin{equation}
    ds ^{2} =a ^{2} ( \eta ) \left\{ - ( 1+2 \Phi ) d\eta ^{2} + \left[ ( 1-2\Phi ) \delta _{ij} + \frac{1}{2} h _{ij} \right] dx ^{i} dx ^{j} \right\},
\end{equation}
where $ \eta $ is the conformal time, $ \Phi $ is the first-order scalar perturbation and $ h _{ij} $ is the induced GWs.
The first-order GWs, the vector perturbations, and the anisotropic stress (\cite{Baumann:2007zm,Weinberg:2003ur,Watanabe:2006qe} showed its effect turns out to be small) are neglected here.
Then the equation of motion for the GWs $ h _{ij} $ can be derived from the Einstein equation straightforwardly.

The Fourier transform of GWs is defined as usual by
\begin{equation}
    h _{ij} ( \bm{x}, \eta) =\int\frac{\dif ^{3} \bm{k}}{( 2\pi )^{3/2}} \me ^{\mi \bm{k}\cdot \bm{x}} [ h _{\bm{k}} (\eta) e _{ij} ( \bm{k} ) + \bar{h} _{\bm{k}} (\eta) \bar{e} _{ij} ( \bm{k} ) ] ,
\end{equation}
where the two time-independent polarization tensors $ e _{ij} (\bm{k})  $ and $ \bar{e} _{ij} (\bm{k})  $ can be written as
\begin{equation}
    e _{ij} (\bm{k}) \equiv \frac{1}{\sqrt{2}} [e _{i} ( \bm{k} )e _{j} ( \bm{k} ) -\bar{e} _{i} ( \bm{k} )\bar{e} _{j} ( \bm{k} )  ] ,
\end{equation}
\begin{equation}
    \bar{e} _{ij} (\bm{k}) \equiv \frac{1}{\sqrt{2}} [e _{i} ( \bm{k} )\bar{e} _{j} ( \bm{k} ) +\bar{e} _{i} ( \bm{k} )e _{j} ( \bm{k} )  ] ,
\end{equation}
and $ e _{i} ( \bm{k} ) $ and $ \bar{e} _{i} ( \bm{k} ) $ are orthonormal basis vectors with respect to $ \bm{k} $.
The source term is defined as
\begin{equation}
        S _{ij} ( \bm{x} ,\eta )  \equiv \ 4\Phi\partial _{i} \partial _{j} \Phi +2 \partial _{i} \Phi \partial _{j} \Phi - \frac{4}{3 ( 1+w ) \mathcal{H} ^{2}} \partial _{i} ( \Phi' + \mathcal{H} \Phi ) \partial _{j} ( \Phi' + \mathcal{H} \Phi),
\end{equation}
where $ w=P/\rho $ is the equation of state parameter of pressure $ P $ and energy density $ \rho $, and $ \mathcal{H} =aH $ is the conformal Hubble parameter, and $ (\dots)' $ denotes a derivative with respect to conformal time $ \eta $.
Then the equation of motion of induced GWs in Fourier space reads
\begin{equation}\label{eq:eof}
    h _{\bm{k}} ''+2 \mathcal{H} h _{\bm{k}} '+ k ^{2} h _{\bm{k}} = S ( \bm{k} ,\eta ) ,
\end{equation}
where
\begin{equation}
        S ( \bm{k} ,\eta ) =-4 e ^{ij} ( \bm{k} )  S _{ij} ( \bm{k} ,\eta )
                           =-4 e ^{ij} ( \bm{k} ) \int \frac{\dif ^{3} \bm{x}}{( 2 \pi ) ^{3/2}} \me ^{- \mi \bm{k} \cdot \bm{x}}  S _{ij} ( \bm{x} ,\eta ).
\end{equation}
This equation of motion can be solved by Green's function method, and the solution is
\begin{equation}
    h _{\bm{k}} ( \eta ) = \frac{1}{a ( \eta )} \int \dif \tilde{\eta} G _{\bm{k}} ( \eta; \tilde{\eta} ) [ a ( \tilde{\eta} ) S ( \bm{k} , \tilde{\eta} ) ] ,
\end{equation}
where the Green's function satisfies
\begin{equation}\label{eq:green}
    G _{\bm{k}} '' + ( k ^{2} - \frac{a''}{a} ) G _{\bm{k}} =\delta ( \eta - \tilde{\eta} ).
\end{equation}

One then splits the Fourier transformation of first-order scalar perturbations $ \Phi _{\bm{k}} ( \eta ) $ into transfer function $ \Phi ( k \eta ) $ and primordial fluctuations $ \phi _{\bm{k}} $,
\begin{equation}
    \Phi _{\bm{k}} ( \eta ) \equiv \Phi ( k \eta ) \phi _{\bm{k}} ,
\end{equation}
so that the transfer function $ \Phi ( k \eta ) $ approaches unity well before the horizon entry.
Now the source term of the equation of motion can be written as
\begin{equation}
    S ( \bm{k} ,\eta ) =\int \frac{\dif ^{3} \tilde{\bm{k}}}{( 2\pi ) ^{3/2}} e ( \bm{k} , \tilde{\bm{k}} ) f ( \bm{k} , \tilde{\bm{k}} , \eta ) \phi _{\bm{k}} \phi _{\bm{k} - \tilde{\bm{k}}} ,
\end{equation}
where
\begin{equation}
    e ( \bm{k} , \tilde{\bm{k}} ) \equiv e ^{ij} ( \bm{k} ) \tilde{k} _{i} \tilde{k} _{j} ,
\end{equation}
\begin{equation}
\begin{aligned}
f ( \bm{k} , \tilde{\bm{k}} , \eta )
=& \frac{8 ( 3w+5 )}{3 ( w+1 )} \Phi (| \tilde{\bm{k}}| \eta ) \Phi ( | \bm{k} - \tilde{\bm{k}}| \eta ) +  \frac{4 ( 3w+1 ) ^{2}}{3 ( w+1 )} \eta ^{2}  \Phi' (| \tilde{\bm{k}}| \eta ) \Phi' ( | \bm{k} - \tilde{\bm{k}}| \eta )\\
&+\frac{8 ( 3w+1 )}{3 ( w+1 )} \eta [  \Phi' (| \tilde{\bm{k}}| \eta ) \Phi ( | \bm{k} - \tilde{\bm{k}}| \eta ) + \Phi (| \tilde{\bm{k}}| \eta ) \Phi' ( | \bm{k} - \tilde{\bm{k}}| \eta ) ].
\end{aligned}
\end{equation}

\subsection{Power spectrum}\label{subsec:spectrum}

The dimensionless power spectrum of GWs is defined by
\begin{equation}\label{eq:power_h}
    \langle h _{\bm{k}} ( \eta ) h _{\bm{l} } ( \eta ) \rangle = \delta ^{( 3 )}( \bm{k} + \bm{l} ) \frac{2 \pi ^{2}}{k ^{3}} \tilde{P} _{h} ( \eta,k ) ,
\end{equation}
and the energy-density spectrum is defined as
\begin{equation}
    \Omega _{\mathrm{GW}} ( \eta ,k ) = \frac{1}{24} \left( \frac{k}{\mathcal{H ( \eta )}} \right) ^{2} \overline{\tilde{P} _{h} ( \eta, k )},
\end{equation}
where the two polarization modes have been summed over, and the overline means oscilllation average or time average\cite{Saito}.
The energy-density spectrum denotes the fraction of the GWs energy density in total energy density per unit logarithmic frequency.

In order to get the observationally relevant quantity $ \Omega _{\mathrm{GW}} ( \eta ,k ) $, one starts with the calculation of the two-point correlation function of $ h _{\bm{k}} $,
\begin{equation}
\langle h _{\bm{k}} ( \eta ) h _{\bm{l} } ( \eta ) \rangle = \left\langle  \frac{1}{a ( \eta )} \int _{\eta _{0}}^{\eta} \dif \tilde{\eta} G _{\bm{k}} ( \eta; \tilde{\eta} ) [ a ( \tilde{\eta} ) S ( \bm{k} , \tilde{\eta} ) ]
\frac{1}{a ( \eta )} \int _{\eta _{0}}^{\eta} \dif \hat{\eta} G _{\bm{l}} ( \eta; \hat{\eta} ) [ a ( \hat{\eta} ) S ( \bm{l} , \hat{\eta} ) ] \right\rangle,
\end{equation}
where the reference time $\eta_0=0$ hereafter.
After defining
\begin{equation}
    I ( \bm{k} , \bm{p} ,\eta ) \equiv \int _{\eta _{0}}^{\eta} \dif \tilde{\eta} \frac{a ( \tilde{\eta} )}{a ( \eta )} G _{\bm{k}} ( \eta ; \tilde{\eta} ) f ( \bm{k} , \bm{p} , \tilde{\eta} ) ,
\end{equation}
one gets
\begin{equation}\label{eq:greenII}
\langle h _{\bm{k}} ( \eta ) h _{\bm{l} } ( \eta ) \rangle = \int \frac{\dif ^{3} p}{( 2\pi ) ^{3/2}} e ( \bm{k} , \bm{p} )  \int \frac{\dif ^{3} q}{( 2\pi ) ^{3/2}} e ( \bm{l} , \bm{q} )
I ( \bm{k} , \bm{p} ,\eta )I ( \bm{l} , \bm{q} ,\eta ) \langle \phi _{\bm{p}} \phi _{\bm{k} -\bm{p}} \phi _{\bm{q}} \phi _{\bm{l} -\bm{q}} \rangle.
\end{equation}

Assuming $ \phi _{\bm{k}} $ is Gaussian, one can utilize the relation of four-point correlator and two-point correlator
\begin{equation}
\langle \phi _{\bm{p}} \phi _{\bm{k} -\bm{p}} \phi _{\bm{q}} \phi _{\bm{l} -\bm{q}} \rangle = \langle \phi _{\bm{p}} \phi _{\bm{k} -\bm{p}} \rangle \langle \phi _{\bm{q}} \phi _{\bm{l} -\bm{q}} \rangle
+\langle \phi _{\bm{p}}  \phi _{\bm{q}}\rangle \langle \phi _{\bm{k} -\bm{p}} \phi _{\bm{l} -\bm{q}} \rangle + \langle \phi _{\bm{p}}  \phi _{\bm{l} -\bm{q}}\rangle \langle \phi _{\bm{k} -\bm{p}}\phi _{\bm{q}}  \rangle,
\end{equation}
and the definition of dimensionless power spectrum of primordial scalar perturbations
\begin{equation}
    \langle \phi _{\bm{k}}  \phi _{\bm{p} }  \rangle = \delta ^{( 3 )}( \bm{k} + \bm{p} ) \frac{2 \pi ^{2}}{k ^{3}} \tilde{P}_{\phi} ( k )  ,
\end{equation}
to simplify \eqref{eq:greenII}.
One obtains
\begin{equation}\label{eq:greenee}
\begin{aligned}
&\langle h _{\bm{k}} ( \eta ) h _{\bm{l} } ( \eta ) \rangle = \delta ^{( 3 )} ( \bm{k} + \bm{l} ) \int \dif ^{3} p \frac{\pi}{2} \frac{1}{| \bm{p} | ^{3} | \bm{k} - \bm{p} | ^{3}} \tilde{P} _{\phi} ( | \bm{p} | ) \tilde{P} _{\phi} ( | \bm{k} - \bm{p} | ) \\
&\times[ e ( \bm{k} , \bm{p} )  e ( -\bm{k} , -\bm{p} )I ( \bm{k} , \bm{p} ,\eta )I ( -\bm{k} , -\bm{p} ,\eta )+e ( \bm{k} , \bm{p} )  e ( -\bm{k} , \bm{p}- \bm{k} )I ( \bm{k} , \bm{p} ,\eta )I ( -\bm{k} , \bm{p}- \bm{k} ,\eta ) ].
\end{aligned}
\end{equation}

Here one introduces three dimensionless variables $ u\equiv| \bm{k} - \bm{p} |/k $, $v\equiv | \bm{p} |/k $ and $ x\equiv k\eta $, and compare \eqref{eq:greenee} with \eqref{eq:power_h}.
Finally one has~\cite{Kohri:2018awv}
\begin{equation}
\tilde{P} _{h} ( \eta, k ) = \frac{1}{4} \int _{0}^{\infty} \dif v \int _{|1-v|}^{1+v} \dif u \left( \frac{4v ^{2} - ( 1+v ^{2} -u ^{2} ) ^{2}}{4uv} \right) ^{2}
\mathcal{I} ^{2} ( u,v,x ) \tilde{P} _{\phi} ( ku ) \tilde{P} _{\phi} ( kv ),
\end{equation}
where
\begin{equation}
    \mathcal{I} ( u,v,x ) \equiv I ( \bm{k} , \bm{p} , \eta ) k ^{2}.
\end{equation}

The time evolution information of power spectrum is contained in the transfer function $ \Phi ( k\eta ) $, which satifies the following constraint equation
\begin{equation}\label{eq:Phi}
    \Phi'' ( k\eta ) + \frac{6 ( 1+w )}{( 1+3w )} \frac{1}{\eta} \Phi' ( k\eta ) +wk ^{2} \Phi ( k\eta ) = 0
\end{equation}
in the absence of entropy perturbations.

\subsection{Radiation era}\label{subsec:rad}

In the RD era, the solution to \eqref{eq:green} and \eqref{eq:Phi} are
\begin{equation}
    kG _{\bm{k}} ( \eta, \tilde{\eta} ) = \sin ( x- \tilde{x} )
\end{equation}
and
\begin{equation}
    \Phi ( k\eta ) = \frac{9}{x ^{2}} \left( \frac{\sin ( x/ \sqrt{3} )}{x/ \sqrt{3}} - \cos ( x/ \sqrt{3} ) \right) ,
\end{equation}
where $ \tilde{x} \equiv k \tilde{\eta} $ and $x \equiv k \eta$.

Since our interset is mainly focused on the GW spectrum observed today, one could take the late-time limit $ \eta \rightarrow \infty $ or $ x \rightarrow \infty $.
Following the methods in\cite{Kohri:2018awv}, one has
\begin{equation}
    \begin{aligned}
        \mathcal{I} ( u,v,x\rightarrow \infty ) = \frac{27}{4 u ^{3} v ^{3} x} ( u ^{2} +v ^{2} -3 )
&\left\{ \sin x \left[ -4uv+( u ^{2} +v ^{2} -3 ) \ln \left|\frac{3- ( u+v ) ^{2}}{3- ( u-v ) ^{2}} \right| \right] \right.\\
&\left.-\cos x \left[ \pi  ( u ^{2} +v ^{2} -3 ) \Theta _{\frac{1}{2}}  ( u+v- \sqrt{3} ) \right] \right\} ,
    \end{aligned}
\end{equation}
where $ \Theta _{s} ( x ) $ is defined as
\begin{equation}
    \Theta _{s} ( x ) = \left\{
    \begin{aligned}
        &1,\quad &x>0\\
        &s ,\quad &x=0\\
        &0, \quad &x<0
    \end{aligned}
    \right.
\end{equation}

In the late-time limit, the energy-density spectrum is given by
\begin{equation}
\Omega _{\mathrm{GW}} ( k ) \equiv \Omega _{\mathrm{GW}} ( \eta \rightarrow \infty, k )
= \frac{1}{24} \int _{0}^{\infty} \dif v \int _{|1-v|}^{1+v} \dif u \mathcal{T} ( u,v ) \tilde{P} _{\phi} ( ku ) \tilde{P} _{\phi} ( kv ) ,
\end{equation}
where
\begin{equation}
    \begin{aligned}
        \mathcal{T} ( u,v )
         &=  \frac{1}{4} \left( \frac{4v ^{2} - ( 1+v ^{2} -u ^{2} ) ^{2}}{4uv} \right) ^{2} \left[  \frac{27}{4 u ^{3} v ^{3}} ( u ^{2} +v ^{2} -3 ) \right] ^{2} \\
          & \times \frac{1}{2} \left\{ \left[ -4uv+( u ^{2} +v ^{2} -3 ) \ln \left|\frac{3- ( u+v ) ^{2}}{3- ( u-v ) ^{2}} \right| \right] ^{2}  +  \left[ \pi  ( u ^{2} +v ^{2} -3 ) \Theta _{\frac{1}{2}}  ( u+v- \sqrt{3} ) \right] ^{2} \right\}.
    \end{aligned}
\end{equation}

\section{Toy model with $\delta$-peak}\label{sec:delta}

We start with a toy model where the dimensionless power spectrum of scalar perturbations exhibits multiple $\delta$-peaks at wave-vectors $k_{*i}$ with proper dimensionless normalization $A_{\phi i}$,
\begin{align}
 \tilde{P}_\phi(k)=\sum_{i=1}^nA_{\phi i}\delta\left(\ln\frac{k}{k_{*i}}\right).
\end{align}
Hereafter $k_{*i}$ is set to meet $0<k_{*1}<k_{*2}<\cdots<k_{*n}$.

\subsection{Single $\delta$-peak}\label{subsec:1delta}

The energy-density spectrum of induced GWs from scalar perturbations with a single $\delta$-peak in power spectrum can be computed directly from
\begin{align}
\Omega_\mathrm{GW}^{1,\delta}(k)=\frac{1}{24}\int_0^\infty\mathrm{d}v\int_{|1-v|}^{1+v}\mathrm{d}u\mathcal{T}(u,v)\tilde{P}_\phi(ku)\tilde{P}_\phi(kv),
\end{align}
which, after noting that $\delta(\ln\tilde{k}u)=\tilde{k}^{-1}\delta(u-\tilde{k}^{-1})$ with $\tilde{k}\equiv k/k_*$, becomes
\begin{align}
\Omega_\mathrm{GW}^{1,\delta}(k)=\frac{1}{24}\int_0^\infty\mathrm{d}v\int_{|1-v|}^{1+v}\mathrm{d}u\mathcal{T}(u,v)\tilde{k}^{-2}\delta(u-\tilde{k}^{-1})\delta(v-\tilde{k}^{-1}),
\end{align}
namely,
\begin{align}
\Omega_\mathrm{GW}^{1,\delta}(k)=\frac{A_{\phi}^2}{24\tilde{k}^2}\mathcal{T}\left(\frac{1}{\tilde{k}},\frac{1}{\tilde{k}}\right)\Theta_0(2-\tilde{k}).
\end{align}

\begin{figure}
\centering
\includegraphics[width=0.48\textwidth]{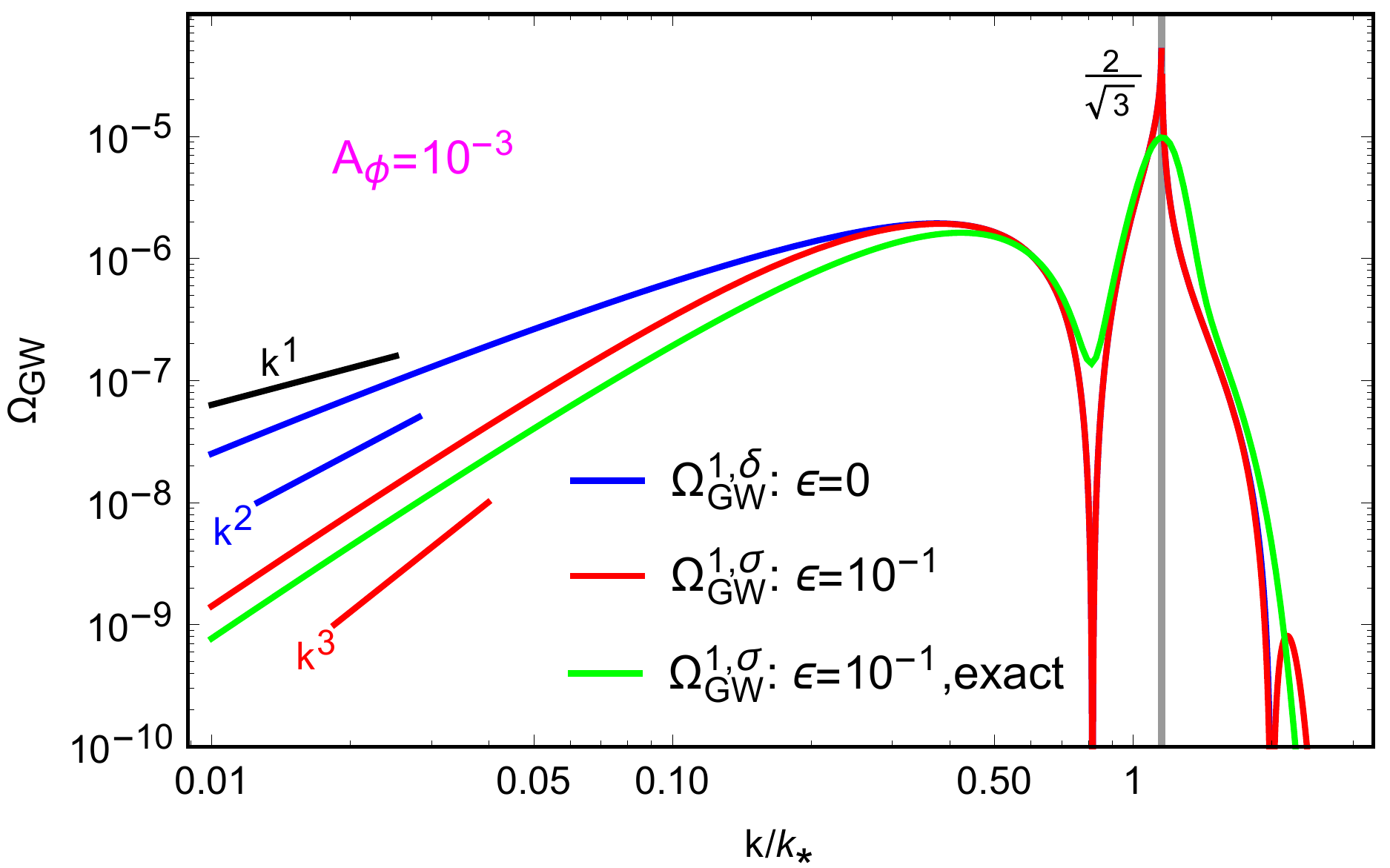}
\includegraphics[width=0.48\textwidth]{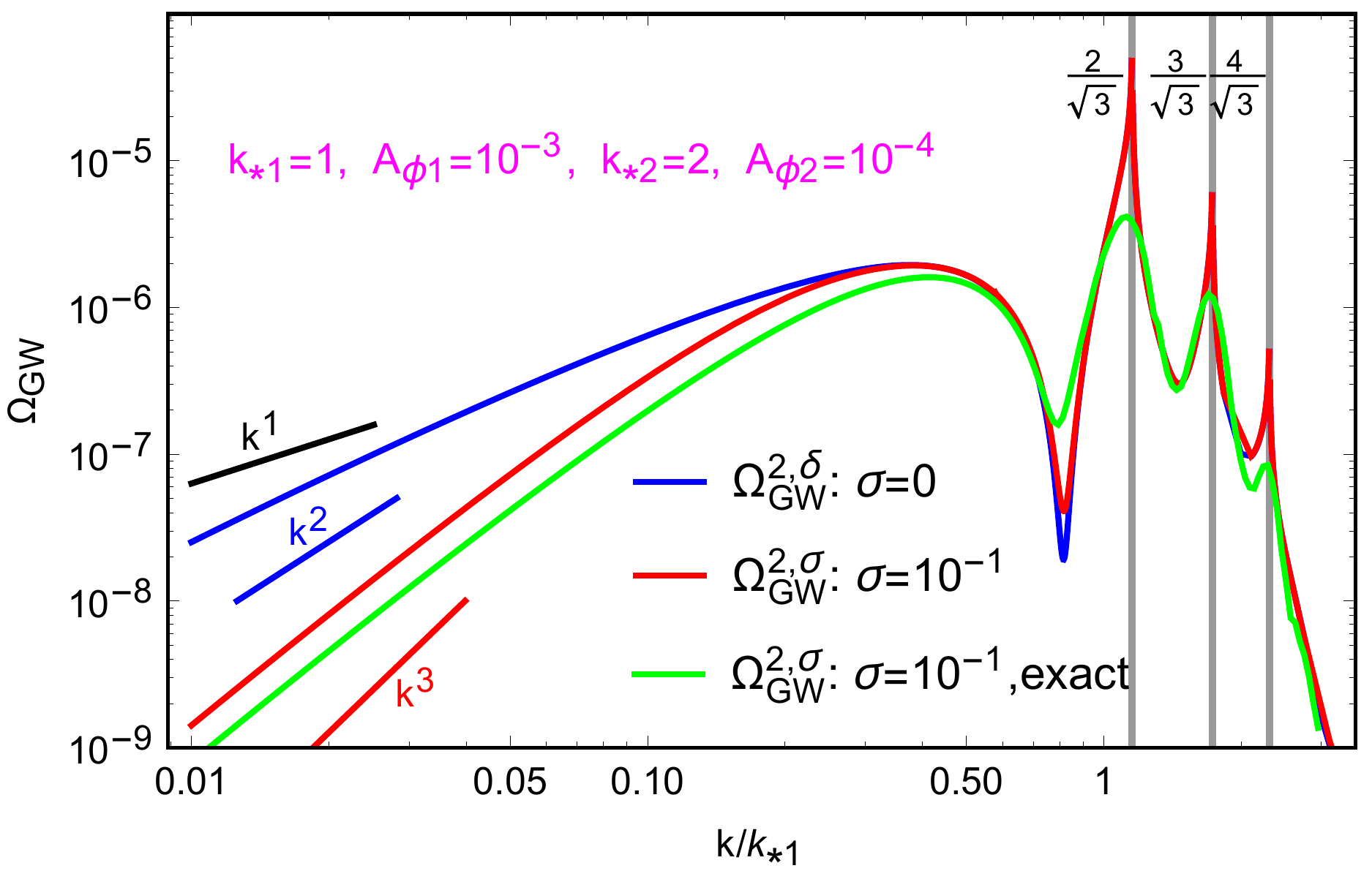}\\
\caption{The energy-density spectrum of induced GWs from primordial scalar perturbations without non-Gaussianity.
\textit{Left}: Induced GWs from primordial scalar perturbations with a single $\delta$-peak (blue), or a single $\sigma$-peak with width $\sigma=10^{-1}k_*$ (red), or a single $\sigma$-peak width $\sigma=10^{-1}k_*$ (green) but without using narrow-width approximation \eqref{eq:narrow}.
\textit{Right}: Induced GWs from primordial scalar perturbations with double $\delta$-peaks (blue), or double $\sigma$-peaks with width $\sigma=10^{-1}k_{*1}$ (red), or double $\sigma$-peaks with width $\sigma=10^{-1}k_{*1}$ (green) but without using narrow-width approximation.
Note that, we are sloppy about the normalization factor $A_\phi$ in all cases, since it only affects the magnitude instead of the position of peaks, thus the vertical axis values are not subjected to any observational reference.}\label{fig:GWs}
\end{figure}

In the left panel of Fig.\ref{fig:GWs},
the energy-density spectrum of induced GWs from scalar perturbations with a single $\delta$-peak in power spectrum is presented with blue line.
As one can see, there is a peak at wave-vector of $k=\frac{2k_*}{\sqrt{3}}$, which can be easily found from the pole of $\mathcal{T}\left(\frac{1}{\tilde{k}},\frac{1}{\tilde{k}}\right)$, namely, $3-\left(\frac{1}{\tilde{k}}+\frac{1}{\tilde{k}}\right)^2=0$ in the logarithmic factor.
It is worth noting that the low-frequency growth of $\Omega_\text{GW}^{1,\delta}$ is around $k^2$, consistent with the observations in\cite{Cai:2018dig,Byrnes:2018txb}.

\subsection{Double $\delta$-peaks}\label{subsec:2delta}

If there are double $\delta$-peaks in the power spectrum of scalar perturbations,
\begin{align}
\tilde{P}_\phi(k)=A_{\phi1}\delta\left(\ln\frac{k}{k_{*1}}\right)+A_{\phi2}\delta\left(\ln\frac{k}{k_{*2}}\right)\equiv\sum_{i=1,2}A_{\phi i}k_{*i}\delta(k-k_{*i}),
\end{align}
the corresponding energy-density spectrum of induced GWs can be computed directly as
\begin{align}
\Omega _\mathrm{GW}^{2,\delta}(k)=&\frac{1}{24}\left[A_{\phi1}^2\frac{k_{*1}^2}{k^2}\mathcal{T}\left(\frac{k_{*1}}{k},\frac{k_{*1}}{k}\right)\Theta_0(2k_{*1}-k)+A_{\phi2}^2\frac{k_{*2}^2}{k^2}\mathcal{T}\left(\frac{k_{*2}}{k},\frac{k_{*2}}{k}\right)\Theta_0(2k_{*2}-k)\right.\nonumber\\
&\left.+2A_{\phi1}A_{\phi2}\frac{k_{*1}k_{*2}}{k^2}\mathcal{T}\left(\frac{k_{*1}}{k},\frac{k_{*2}}{k}\right)\Theta_0(k_{*1}+k_{*2}-k)\Theta_0(k-|k_{*1}-k_{*2}|)\right].
\end{align}

In the right panel of Fig.\ref{fig:GWs}, the energy-density spectrum of induced GWs from scalar perturbations with double $\delta$-peaks in power spectrum is presented with blue solid line.
As one can see, there is a  triple-peak structure at around $k_{*1}$, whose wave-vectors can be inferred from the pole of $\mathcal{T}(u,v)$,
\begin{align}
3-(u+v)^2=0,\quad 4v^2-(1+v^2-u^2)^2\neq0,\quad u^2+v^2-3\neq0,
\end{align}
i.e.,
\begin{align}
u+v=\sqrt{3},\quad \left\{\begin{matrix}
u\neq0\\
v\neq\sqrt{3}
\end{matrix}\right.,\quad \left\{\begin{matrix}
u\neq\sqrt{3}\\
v\neq0
\end{matrix}\right.,\quad \left\{\begin{matrix}
u\neq\frac12(\sqrt{3}\pm1)\\
v\neq\frac12(\sqrt{3}\mp1)
\end{matrix}\right.
\end{align}
If one further requires $v>0$ and $|1-v|<u<1+v$, then the pole of $\mathcal{T}(u,v)$ would be simply from the condition $u+v=\sqrt{3}$.
Therefore, for our double $\delta$-peaks, the three poles of
\begin{align}
\mathcal{T}\left(\frac{k_{*1}}{k},\frac{k_{*1}}{k}\right), \quad
\mathcal{T}\left(\frac{k_{*1}}{k},\frac{k_{*2}}{k}\right), \quad
\mathcal{T}\left(\frac{k_{*2}}{k},\frac{k_{*2}}{k}\right)
\end{align}
are given by the conditions
\begin{align}
\frac{k_{*1}}{k}+\frac{k_{*1}}{k}=\sqrt{3},\quad
\frac{k_{*1}}{k}+\frac{k_{*2}}{k}=\sqrt{3},\quad
\frac{k_{*2}}{k}+\frac{k_{*2}}{k}=\sqrt{3},
\end{align}
respectively, namely,
\begin{align}
k=\frac{1}{\sqrt{3}}(k_{*1}+k_{*1}),\quad
\frac{1}{\sqrt{3}}(k_{*1}+k_{*2}),\quad
\frac{1}{\sqrt{3}}(k_{*2}+k_{*2}).
\end{align}
A special case is that $k_{*2}=2k_{*1}$ for our double $\delta$-peaks in scalar perturbations.
The energy-density spectrum $\Omega _\mathrm{GW}^{2,\delta}(k)$ of induced GWs would produce the triple-peak structure
\begin{align}
\frac{k}{k_{*1}}=\frac{2}{\sqrt{3}}, \frac{3}{\sqrt{3}}, \frac{4}{\sqrt{3}},
\end{align}
which has the similar structure for GW energy-density spectrum induced by sclar perturbations with a single narrow peak and primordial non-Gaussianities\cite{Cai:2018dig}.

\subsection{Multiple $\delta$-peaks}\label{subsec:ndelta}

The general case of multiple($n$) $\delta$-peaks in the scalar perturbations
\begin{align}
\tilde{P}_\phi(k)=\sum_{i=1}^nA_{\phi i}k_{*i}\delta(k-k_{*i})
\end{align}
goes parallel to the double $\delta$-peaks, and the corresponding energy-density spectrum of induced GWs reads
\begin{align}
\Omega_\mathrm{GW}^{n,\delta}(k)=\frac{1}{24}\sum_{i,j=1}^nA_{\phi i}A_{\phi j}k_{*i}k_{*j}\int_0^\infty\mathrm{d}v\int_{|1-v|}^{1+v}\mathrm{d}u\mathcal{T}(u,v)\delta(ku-k_{*i})\delta(kv-k_{*j}),
\end{align}
namely,
\begin{align}\label{eq:GWn-delta}
\Omega_\mathrm{GW}^{n,\delta}(k)=\frac{1}{24}\sum_{i,j=1}^nA_{\phi i}A_{\phi j}\frac{k_{*i}k_{*j}}{k^2}\mathcal{T}\left(\frac{k_{*i}}{k},\frac{k_{*j}}{k}\right)\Theta_0(k_{*i}+k_{*j}-k)\Theta_0(k-|k_{*i}-k_{*j}|),
\end{align}
with $C_{n+1}^2$ peaks given by
\begin{align}
k_{ij}=\frac{1}{\sqrt{3}}(k_{*i}+k_{*j}).
\end{align}

It is worth noting that, there are at most $C_{n+1}^2$ and at least $n$ peaks, because some of $k_{ij}$ could be identical for the combination $(k_{*i}+k_{*j})/\sqrt{3}$, and some of peaks at $k_{ij}$ vanish due to $\Theta_0$ if $k_{ij}\notin \left(|k_{*i}-k_{*j}|,k_{*i}+k_{*j}\right)$.
The obtained multiple-peak structure can be understood as resonant amplification, which can be easily seen from the equation of motion \eqref{eq:eof} of form
\begin{align}
h _{\bm{k}} ''+k ^{2} h _{\bm{k}} \sim S ( \bm{k} ,\eta ) \sim \sum_{i,j=1}^{n} \sin(\frac{k_{*i}}{\sqrt{3}}\eta) \sin(\frac{k_{*j}}{\sqrt{3}}\eta),
\end{align}
here $h_{\bm{k}}$ is resonantly amplified when
\begin{align}
k=\frac{+k_{*i}+k_{*j}}{\sqrt{3}},\frac{+k_{*i}-k_{*j}}{\sqrt{3}},\frac{-k_{*i}+k_{*j}}{\sqrt{3}},\frac{-k_{*i}-k_{*j}}{\sqrt{3}}.
\end{align}
For convenience, we introduce here the dubbed wave-vector factor
\begin{align}\label{eq:F_delta}
\mathcal{F}^\delta(k;k_{*i},k_{*j})\equiv\Theta_0(k_{*i}+k_{*j}-k)\Theta_0(k-|k_{*i}-k_{*j}|),
\end{align}
that will be interpreted as result of momentum conservation in the next section.

\section{Realistic model with $\sigma$-peak}\label{sec:sigma}

$\delta$-function peak has infinitesimal width which seems not natural, as usual inflation models predict primordial scalar perturbations with finite peaks.
Therefore, we turn to more realistic Gaussian peaks with finite width $\sigma$, dubbed $\sigma$-peak model, which can be parameterized as
\begin{align}
P_\phi(k)=\frac{A_\phi}{(2\pi)^{3/2}2\sigma k_*^2}\mathrm{e}^{-\frac{(k-k_*)^2}{2\sigma^2}}.
\end{align}
Here, we require that the width of the peak is narrow, i.e. $k_*\gg\sigma$, and $A_\phi$ is a dimensionless amplitude,
\begin{align}
\int\mathrm{d}^3kP_\phi(k)=\int\mathrm{d}k4\pi k^2P_\phi(k)\approx4\pi k_*^2\int\mathrm{d}kP_\phi(k)=A_\phi
\end{align}
whose precise value is not of our concern as for our purpose to show.

\subsection{Single $\sigma$-peak}\label{subsec:1sigma}

Using the rescaled dimensionless parameters $\tilde{k}\equiv k/k_*$ and $\epsilon\equiv\sigma/k_*\ll1$, one can also rewrite the primordial scalar power spectrum as a dimensionless form,
\begin{align}
\tilde{P}_\phi(\tilde{k})\equiv4\pi k^3P_\phi(k)=\frac{A_\phi\tilde{k}^3}{\sqrt{2\pi}\epsilon}\mathrm{e}^{-\frac{(\tilde{k}-1)^2}{2\epsilon^2}}.
\end{align}
The energy-density spectrum of induced GWs from scalar perturbations with a single $\sigma$-peak in power spectrum can thus be computed as
\begin{align}
\Omega_\mathrm{GW}^{1,\sigma}(\tilde{k})&=\frac{1}{24}\int_0^\infty\mathrm{d}v\int_{|1-v|}^{1+v}\mathrm{d}u\mathcal{T}(u,v)\tilde{P}_\phi(\tilde{k}u)\tilde{P}_\phi(\tilde{k}v)\nonumber\\
&=\frac{A_\phi^2\tilde{k}^6}{24(2\pi)\epsilon^2}\int_0^\infty\mathrm{d}v\int_{|1-v|}^{1+v}\mathrm{d}u\mathcal{T}(u,v)u^3v^3\mathrm{e}^{-\frac{(\tilde{k}u-1)^2+(\tilde{k}v-1)^2}{2\epsilon^2}}.
\end{align}

For a sufficiently narrow width of $\sigma$-peak ($\epsilon \ll 1$), the energy-density spectrum of induced GWs can be approximated (narrow-width approximation) as
\begin{align}\label{eq:narrow}
\Omega_\mathrm{GW}^{1,\sigma}(\tilde{k})\approx\frac{A_\phi^2\tilde{k}^6}{24(2\pi)\epsilon^2}\frac{\mathcal{T}\left(\frac{1}{\tilde{k}},\frac{1}{\tilde{k}}\right)}{\tilde{k}^3\tilde{k}^3}\int_0^\infty\mathrm{d}v\int_{|1-v|}^{1+v}\mathrm{d}u\,\mathrm{e}^{-\frac{(\tilde{k}u-1)^2+(\tilde{k}v-1)^2}{2\epsilon^2}}.
\end{align}
After turning to the new variables $s=u+v$ and $t=u-v$, the energy-density spectrum of induced GWs
\begin{align}
\Omega_\mathrm{GW}^{1,\sigma}(\tilde{k})=\frac{A_\phi^2}{24(2\pi)\epsilon^2}\mathcal{T}\left(\frac{1}{\tilde{k}},\frac{1}{\tilde{k}}\right)\int_1^\infty\mathrm{d}s\int_{-1}^{1}\mathrm{d}t\,\frac12\mathrm{e}^{-\frac{\tilde{k}^2t^2+(\tilde{k}s-2)^2}{4\epsilon^2}}
\end{align}
can be integrated analytically as
\begin{align}
\Omega_\mathrm{GW}^{1,\sigma}(\tilde{k})=\frac{A_\phi^2}{24(2\pi)\epsilon^2}\mathcal{T}\left(\frac{1}{\tilde{k}},\frac{1}{\tilde{k}}\right)\pi\epsilon^2\tilde{k}^{-2}\mathrm{erf}\left(\frac{\tilde{k}}{2\epsilon}\right)\left[1+\mathrm{erf}\left(\frac{2-\tilde{k}}{2\epsilon}\right)\right],
\end{align}
namely,
\begin{align}\label{eq:GW1-sigma}
\Omega_\mathrm{GW}^{1,\sigma}(\tilde{k})=\frac{A_\phi^2}{24\tilde{k}^2}\mathcal{T}\left(\frac{1}{\tilde{k}},\frac{1}{\tilde{k}}\right)\frac12\mathrm{erf}\left(\frac{\tilde{k}}{2\epsilon}\right)\left[1+\mathrm{erf}\left(\frac{2-\tilde{k}}{2\epsilon}\right)\right].
\end{align}

In the vanishing width limit ($\epsilon\rightarrow0$) of $\sigma$-peak,
\begin{align}
\mathrm{erf}\left(\frac{\tilde{k}}{2\epsilon}\right)\rightarrow1,\quad\frac12\left[1+\mathrm{erf}\left(\frac{2-\tilde{k}}{2\epsilon}\right)\right]\rightarrow\Theta_\frac12(2-\tilde{k}),
\end{align}
the energy-density spectrum of induced GWs \eqref{eq:GW1-sigma} recovers the result of single $\delta$-peak as expected,
\begin{align}
\lim_{\sigma\rightarrow0}\Omega_\mathrm{GW}^{1,\sigma}(\tilde{k})=\frac{A_\phi^2}{24\tilde{k}^2}\mathcal{T}\left(\frac{1}{\tilde{k}},\frac{1}{\tilde{k}}\right)\Theta_\frac12(2-\tilde{k})=\Omega_\mathrm{GW}^{1,\delta}(\tilde{k}).
\end{align}
Note that the difference of $\Theta_0(x)$ and $\Theta_\frac12(x)$ does not affect the result since $\mathcal{T}(\frac12,\frac12)=0$.

In the left panel of Fig.\ref{fig:GWs}, the energy-density spectrum of induced GWs from a single $\sigma$-peak with width $\sigma=10^{-1}k_*$ in scalar perturbations is presented as a red solid line, which manifests exactly the same peak structure around the scale $k=\frac{2k_*}{\sqrt{3}}$ as in the case of single $\delta$-peak.
As a comparison, we also present with green solid line the induced GWs from a single $\sigma$-peak with the same width $\sigma=10^{-1}k_*$ in scalar perturbations but without using the narrow-width approximation \eqref{eq:narrow}.
As one can see, the small-scale peak position remains unchanged but becomes less cuspy, and the slightly suppressed large-scale growth behaves exactly the same $k^3$-law as found in the case with non-Gaussianity\cite{Cai:2018dig}.

\subsection{Multiple $\sigma$-peaks}\label{subsec:nsigma}

Now we generalize above derivation into the case of multiple($n$) $\sigma$-peaks with $n\geq2$.
The dimensionless power spectrum of scalar perturbation is defined by
\begin{align}
\tilde{P}_{\phi}(k)=\sum\limits_{i=1}^n\frac{A_{\phi i}k_{*i}}{\sqrt{2\pi}\sigma}\mathrm{e}^{-\frac{(k-k_{*i})^2}{2\sigma^2}},
\end{align}
and the corresponding energy-density spectrum of induced GWs from scalar perturbations with such multiple $\sigma$-peaks in power spectrum reads
\begin{align}
\Omega_\mathrm{GW}^{n,\sigma}(k)&=\frac{1}{24}\int_0^\infty\mathrm{d}v\int_{|1-v|}^{1+v}\mathrm{d}u\mathcal{T}(u,v)\tilde{P}_\phi(k u)\tilde{P}_\phi(k v)\nonumber\\
&=\frac{1}{24}\sum\limits_{i,j=1}^nA_{\phi i}A_{\phi j}\frac{k_{*i}k_{*j}}{2\pi\sigma^2}\int_0^\infty\mathrm{d}v\int_{|1-v|}^{1+v}\mathrm{d}u\mathcal{T}(u,v)\mathrm{e}^{-\frac{(ku-k_{*i})^2}{2\sigma^2}-\frac{(kv-k_{*j})^2}{2\sigma^2}}.
\end{align}

\begin{figure}
\centering
\includegraphics[width=0.49\textwidth]{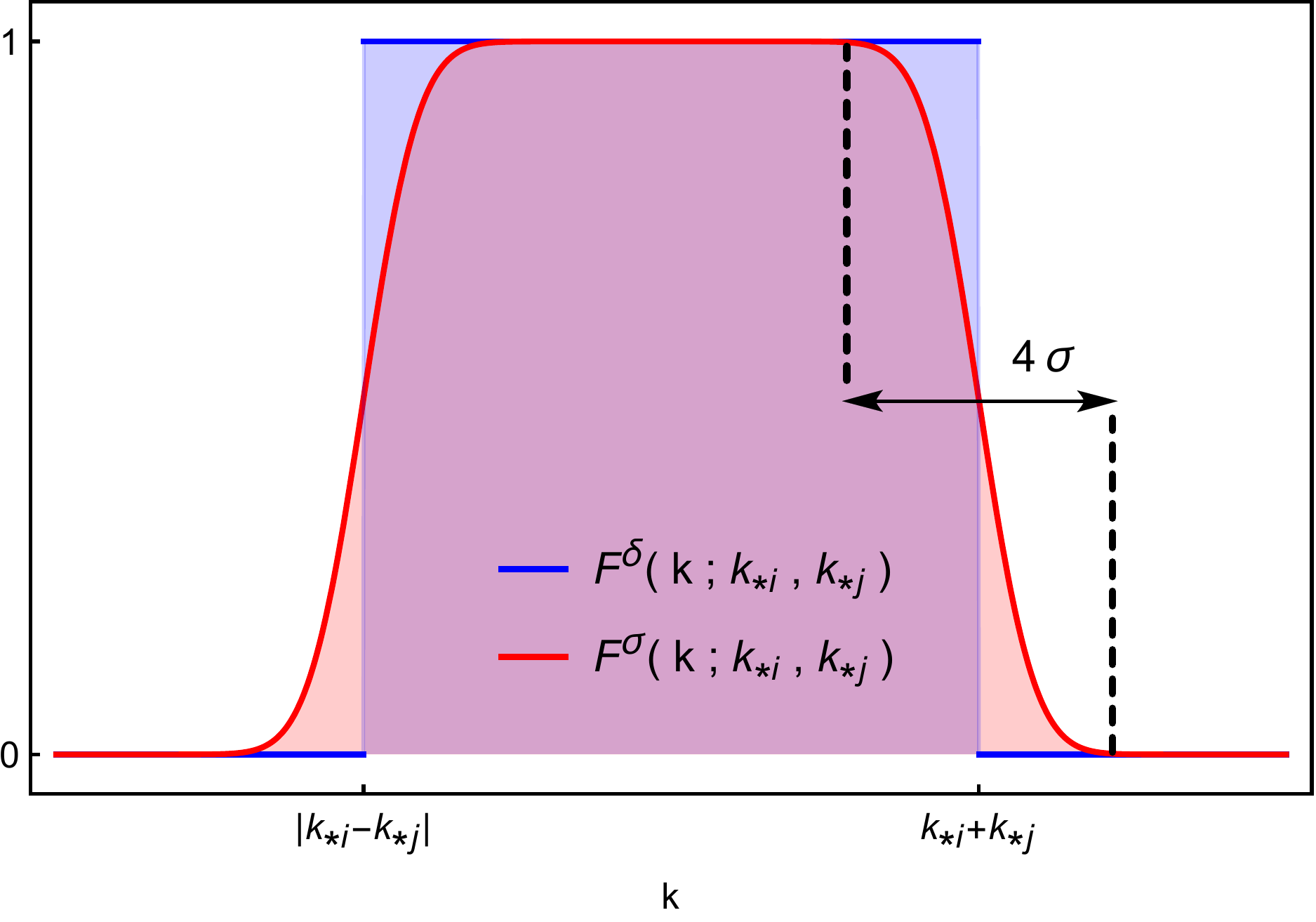}
\caption{Illustration of the wave-vector factors $\mathcal{F}^\delta(k;k_{*i},k_{*j})$ and $\mathcal{F}^\sigma(k;k_{*i},k_{*j})$ of $\delta$-peak (blue) and of $\sigma$-peak (red) with respect to the norm of wave-vector $k$.}\label{fig:factor}
\end{figure}

For a sufficiently narrow width ($k_{*i}\gg\sigma$) and sufficiently distant ($|k_{*i}-k_{*j}|\gg\sigma$)  $\sigma$-peak, the energy-density spectrum of induced GWs can be approximated (narrow-width approximation) as
\begin{align}\label{eq:n_narrow}
\Omega_\mathrm{GW}^{n,\sigma}(k)\approx\frac{1}{24}\sum\limits_{i,j=1}^nA_{\phi i}A_{\phi j}\frac{k_{*i}k_{*j}}{2\pi\sigma^2}\mathcal{T}\left(\frac{k_{*i}}{k},\frac{k_{*j}}{k}\right)\int_0^\infty\mathrm{d}v\int_{|1-v|}^{1+v}\mathrm{d}u\,\mathrm{e}^{-\frac{(ku-k_{*i})^2}{2\sigma^2}-\frac{(kv-k_{*j})^2}{2\sigma^2}}.
\end{align}
After turning to the new variables $s=u+v$ and $t=u-v$, the GWs energy-density spectrum
\begin{align}
\Omega_\mathrm{GW}^{n,\sigma}(k)=\frac{1}{24}\sum\limits_{i,j=1}^nA_{\phi i}A_{\phi j}\frac{k_{*i}k_{*j}}{2\pi\sigma^2}\mathcal{T}\left(\frac{k_{*i}}{k},\frac{k_{*j}}{k}\right)\int_{1}^\infty\mathrm{d}s\int_{-1}^{1}\mathrm{d}t\,\frac12\mathrm{e}^{-\frac{(ku-k_{*i})^2}{2\sigma^2}-\frac{(kv-k_{*j})^2}{2\sigma^2}}
\end{align}
can be integrated analytically as
\begin{align}
\Omega_\mathrm{GW}^{n,\sigma}(k)&=\frac{1}{24}\sum\limits_{i,j=1}^nA_{\phi i}A_{\phi j}\frac{k_{*i}k_{*j}}{2\pi\sigma^2}\mathcal{T}\left(\frac{k_{*i}}{k},\frac{k_{*j}}{k}\right)\frac{\pi\sigma^2}{2k^2}\nonumber\\
&\times\left[\mathrm{erf}\left(\frac{k-(k_{*i}-k_{*j})}{2\sigma}\right)+\mathrm{erf}\left(\frac{k+(k_{*i}-k_{*j})}{2\sigma}\right)\right]\left[1+\mathrm{erf}\left(\frac{k_{*i}+k_{*j}-k}{2\sigma}\right)\right],
\end{align}
namely
\begin{align}\label{eq:GWn-sigma}
\Omega_\mathrm{GW}^{n,\sigma}(k)=\frac{1}{24}\sum\limits_{i,j=1}^nA_{\phi i}A_{\phi j}\frac{k_{*i}k_{*j}}{k^2}\mathcal{T}\left(\frac{k_{*i}}{k},\frac{k_{*j}}{k}\right)\mathcal{F}^\sigma(k;k_{*i},k_{*j}),
\end{align}
where we have introduced the wave-vector factor of $\sigma$-peak
\begin{align}\label{eq:F_sigma}
\mathcal{F}^\sigma(k;k_{*i},k_{*j})\equiv\frac14\left[\mathrm{erf}\left(\frac{k-|k_{*i}-k_{*j}|}{2\sigma}\right)+\mathrm{erf}\left(\frac{k+|k_{*i}-k_{*j}|}{2\sigma}\right)\right]\left[1+\mathrm{erf}\left(\frac{k_{*i}+k_{*j}-k}{2\sigma}\right)\right]
\end{align}
showed in the Fig.\ref{fig:factor} along with previously defined wave-vector factor $\mathcal{F}^\delta(k;k_{*i},k_{*j})$ of $\delta$-peak.

In the right panel of Fig.\ref{fig:GWs}, the energy-density spectrum of induced GWs from scalar perturbations with double $\sigma$-peaks in power spectrum is presented as red solid line.
The wave-vectors of peak position are set at $k_{*1}=1$ and $k_{*2}=2$ with width $\sigma=10^{-1}k_{*1}$ and amplitude $A_{\phi1}=10^{-3}$ and $A_{\phi2}=10^{-4}$, respectively.
As one can see, the peak structure of induced GWs is exactly the same as in the case of double $\delta$-peaks, which also reproduces exactly the same triple-peak structure at the scales $k/k_{*1}=2/\sqrt{3}, 3/\sqrt{3}, 4/\sqrt{3}$ observed recently in the case of single $\sigma$-peak in scalar perturbation with non-Gaussianity.
We also present with green solid line the induced GWs from double $\sigma$-peaks with the same width $\sigma=10^{-1}k_{*1}$ in scalar perturbations but without using the narrow-width approximation \eqref{eq:n_narrow}.
As one can see, the small-scale peak positions remain unchanged but becomes less cuspy, and the slightly suprressed large-scale growth behaves exactly the same $k^3$-law as found in\cite{Cai:2018dig}.

\subsection{Wave-vector factor}\label{subsec:triangle}

In the vanishing width limit ($\sigma\rightarrow0$) of $\sigma$-peak, the wave-vector factor $\mathcal{F}^\sigma(k;k_{*i},k_{*j})$ can be obviously reduced to the previously defined wave-vector factor $\mathcal{F}^\delta(k;k_{*i},k_{*j})$, both of which describe some kind of constraint condition on $k$.
We will show below that such a constraint of wave-vector factor comes from momentum conservation.

Since the GWs ($h_{\bm{k}}$) are induced by the scalar perturbations and there are two scalar perturbations $\phi_{\bm{k}_i}$ and $\phi_{\bm{k}_j}$ in the source term of the equation of motion \eqref{eq:eof}, the momentum conservation requires
\begin{align}
\bm{k}=\bm{k}_i+\bm{k}_j,
\end{align}
namely
\begin{align}
\underline{k}_{ij} \le k \le \overline{k}_{ij},
\end{align}
where
\begin{align}
\underline{k}_{ij}\equiv\min|\bm{k}_i+\bm{k}_j|=|k_i-k_j|,\qquad \overline{k}_{ij}\equiv\max|\bm{k}_i+\bm{k}_j|=k_i+k_j.
\end{align}
For dimensionless power spectrum of scalar perturbations with multiple $\sigma$-peaks,
\begin{align}
\tilde{P}_{\phi}(k)=\sum\limits_{i=1}^n\frac{A_{\phi i}k_{*i}}{\sqrt{2\pi}\sigma}\mathrm{e}^{-\frac{(k-k_{*i})^2}{2\sigma^2}},
\end{align}
the probability density of scalar perturbations $\phi$ with momentum $k_i$ is assumed as
\begin{align}
    p ^{\sigma}(k_i)=\frac{1}{\sqrt{2\pi}\sigma}\mathrm{e}^{-\frac{(k_i-k_{*i})^2}{2\sigma^2}},
\end{align}
and then the probability of GWs $ h _{ij} $ with momentum $ k $ can be given by
\begin{equation}
    \mathcal{P} ^{\sigma} ( \underline{k}_{ij} \le k \le \overline{k}_{ij}) = \mathcal{P} ^{\sigma} ( \underline{k}_{ij} \le k) \mathcal{P} ^{\sigma} (\overline{k}_{ij}\ge k).
\end{equation}
Setting $ \mu =k _{i} + k _{j} $ and $ \nu=k _{i} - k _{j} $, we have
\begin{align}
    \mathcal{P} ^{\sigma}( \underline{k}_{ij} \le k)
    &=\int _{-\infty}^{\infty} \dif\mu \int _{-k}^{k} \dif \nu \frac{1}{2} p^\sigma(k_i)p^\sigma(k_j)\nonumber\\
    &= \frac{1}{2} \left[ \mathrm{erf} \left( \frac{k - |k _{*i} -k _{*j}|}{2\sigma} \right)+ \mathrm{erf} \left( \frac{k + |k _{*i} -k _{*j}|}{2\sigma}\right)  \right] ;\\
    \mathcal{P} ^{\sigma} ( \overline{k}_{ij} \ge k)
    &=\int _{k}^{\infty} \dif\mu \int _{-\infty}^{\infty} \dif \nu \frac{1}{2} p^\sigma(k_i)p^\sigma(k_j)\nonumber\\
    &= \frac{1}{2} \left[ 1 + \mathrm{erf} \left( \frac{k _{*i} +k _{*j} -k}{2\sigma}\right)  \right] ,
\end{align}
which immediately gives rise to
\begin{equation}
    \mathcal{F}^\sigma(k;k_{*i},k_{*j}) = \mathcal{P} ^{\sigma} ( \underline{k}_{ij} \le k \le \overline{k}_{ij} ).
\end{equation}
Similarly, for scalar perturbation with multiple $\delta$-peaks, we also have
\begin{equation}
    \mathcal{F}^\delta(k;k_{*i},k_{*j}) = \mathcal{P} ^{\delta} ( \underline{k}_{ij} \le k \le \overline{k}_{ij} ).
\end{equation}
Therefore, the wave-vector factor of both $\delta$-peak and $\sigma$-peak can be interpreted as the probability of having wave-vector of GWs to obey the momentum conservation.

\section{Induced GWs from non-Gaussian scalar perturbations}\label{sec:NGGWs}

The induced GWs from non-Gaussian scalar perturbations with a single peak was studied in\cite{Cai:2018dig}.
In this section, we take an analytical investigation of GWs induced by multiple peaks of scalar perturbations with non-Gaussianity.

\subsection{Energy-density spectrum}\label{subsec:NGGWs}

In order to get the energy-density spectrum of induced GWs, we start from \eqref{eq:greenII}.
For a non-Gaussian $ \phi _{\bm{k}}^{\mathrm{NG}} $, the four-point correlator can be written as
\begin{align}
\langle \phi^\mathrm{NG} _{\bm{p}} \phi^\mathrm{NG} _{\bm{k} -\bm{p}} \phi^\mathrm{NG} _{\bm{q}} \phi^\mathrm{NG} _{\bm{l} -\bm{q}} \rangle =
&\langle \phi^\mathrm{NG} _{\bm{p}} \phi^\mathrm{NG} _{\bm{k} -\bm{p}} \rangle \langle \phi^\mathrm{NG} _{\bm{q}} \phi^\mathrm{NG} _{\bm{l} -\bm{q}} \rangle
+\langle \phi^\mathrm{NG} _{\bm{p}}  \phi^\mathrm{NG} _{\bm{q}}\rangle \langle \phi^\mathrm{NG} _{\bm{k} -\bm{p}} \phi^\mathrm{NG} _{\bm{l} -\bm{q}} \rangle \nonumber\\
&+ \langle \phi^\mathrm{NG} _{\bm{p}}  \phi^\mathrm{NG} _{\bm{l} -\bm{q}}\rangle \langle \phi^\mathrm{NG} _{\bm{k} -\bm{p}}\phi^\mathrm{NG} _{\bm{q}}  \rangle
+ \langle \phi^\mathrm{NG} _{\bm{p}} \phi^\mathrm{NG} _{\bm{k} -\bm{p}} \phi^\mathrm{NG} _{\bm{q}} \phi^\mathrm{NG} _{\bm{l} -\bm{q}} \rangle _{c},
\end{align}
since the one-point correlator is zero, where $ \langle \dots \rangle _{c} $ is the connected moment.
Note that\cite{Seery.JoCaAP.2007.jan,Byrnes.PRD.2006.11}
\begin{equation}
    \langle \phi^\mathrm{NG} _{\bm{p}} \phi^\mathrm{NG} _{\bm{k} -\bm{p}} \phi^\mathrm{NG} _{\bm{q}} \phi^\mathrm{NG} _{\bm{l} -\bm{q}} \rangle _{c}= ( 2\pi ) ^{3} \delta ^{( 3 )} ( \bm{k} + \bm{l} ) T _{\phi^\mathrm{NG}} ,
\end{equation}
where $T _{\phi^\mathrm{NG}}$ is a function of different momenta.
In comparison to correlators like $\langle\phi\phi\rangle\langle\phi\phi\rangle$, the key point of $\langle\phi\phi\phi\phi\rangle_c$ is that, it contains only one overall $\delta$-function of $\delta^{(3)}(\bm{k}+\bm{l})$, which gives
\begin{equation}
\int \frac{\dif ^{3} p}{( 2\pi ) ^{3/2}} e ( \bm{k} , \bm{p} )  \int \frac{\dif ^{3} q}{( 2\pi ) ^{3/2}} e ( \bm{l} , \bm{q} )
    I ( \bm{k} , \bm{p} ,\eta )I ( \bm{l} , \bm{q} ,\eta ) \langle \phi^\mathrm{NG} _{\bm{p}} \phi^\mathrm{NG} _{\bm{k} -\bm{p}} \phi^\mathrm{NG} _{\bm{q}} \phi^\mathrm{NG} _{\bm{l} -\bm{q}} \rangle _{c}=0,
\end{equation}
since now the two integrals on the azimuth angles are independent thus gives zero result.
Therefore, the energy-density spectrum of induced GWs can be computed by
\begin{align}\label{eq:NGGW}
\Omega_\mathrm{GW}(k)=\frac{1}{24}\int_0^\infty\mathrm{d}v\int_{|1-v|}^{1+v}\mathrm{d}u\mathcal{T}(u,v)\tilde{P}_\phi^\mathrm{NG}(ku)\tilde{P}_\phi^\mathrm{NG}(kv),
\end{align}
where the (dimensionless) power spectrum for a non-Gaussian $\phi_{\bm{k}}^{\mathrm{NG}}$ is defined as
\begin{align}
	\langle\phi^\mathrm{NG}_{\bm{k}}\phi^\mathrm{NG}_{\bm{p}}\rangle
	=(2\pi)^3\delta^{(3)}(\bm{k}+\bm{p})P^\mathrm{NG}_\phi(k)
	=\delta^{(3)}(\bm{k}+\bm{p})\frac{2\pi^2}{k^3}\tilde{P}_\phi^\mathrm{NG}(k).
\end{align}

\subsection{Multiple $\delta$-peaks}\label{subsec:NGndelta}

For simplicity, we consider non-Gaussian scalar perturbations with local type non-Gaussianity
\begin{align}
	\phi^\mathrm{NG}(\bm{x})=\phi(\bm{x})+f_\mathrm{NL}\left[\phi^2(\bm{x})-\langle\phi^2(\bm{x})\rangle\right],
\end{align}
where $ \phi ( \bm{x} ) $ is the Gaussian perturbation.
We get the corresponding (dimensionless) power spectrum
\begin{align}
	\langle\phi^\mathrm{NG}_{\bm{k}}\phi^\mathrm{NG}_{\bm{p}}\rangle
	&=(2\pi)^3\delta^{(3)}(\bm{k}+\bm{p})P^\mathrm{NG}_\phi(k)
	=\delta^{(3)}(\bm{k}+\bm{p})\frac{2\pi^2}{k^3}\tilde{P}_\phi^\mathrm{NG}(k),\\
	&=(2\pi)^3\delta^{(3)}(\bm{k}+\bm{p})\left[P_\phi(k)+2f_\mathrm{NL}^2\int\mathrm{d}^3lP_\phi(|\bm{l}|)P_\phi(|\bm{k}-\bm{l}|)\right].
\end{align}
Then the dimensionless power spectrum reads
\begin{align}
	\tilde{P}_\phi^\mathrm{NG}(k)=\tilde{P}_\phi(k)+\frac{k^3}{2\pi}f_\mathrm{NL}^2\int\mathrm{d}^3l\frac{1}{|\bm{l}|^3|\bm{k}-\bm{l}|^3}\tilde{P}_\phi(|\bm{l}|)\tilde{P}_\phi(|\bm{k}-\bm{l}|).
\end{align}
After introducing new variables $u=|\bm{k}-\bm{l}|/k$ and $v=|\bm{l}|/k$, one finds the dimensionless power spectrum of form
\begin{align}\label{eq:NGP}
	\tilde{P}_\phi^\mathrm{NG}(k)=\tilde{P}_\phi(k)+f_\mathrm{NL}^2\int_0^\infty\mathrm{d}v\int_{|1-v|}^{1+v}\mathrm{d}u\frac{1}{u^2v^2}\tilde{P}_\phi(ku)\tilde{P}_\phi(kv).
\end{align}

Consider the case of multiple($n$) $\delta$-peaks in the power spectrum of a non-Gaussian scalar perturbation,
\begin{align}
\tilde{P}_\phi(k)=\sum_{i=1}^nA_{\phi i}\delta\left(\ln\frac{k}{k_{*i}}\right)=\sum_{i=1}^nA_{\phi i}k_{*i}\delta(k-k_{*i}),
\end{align}
then the power spectrum \eqref{eq:NGP} with non-Gaussianity reads
\begin{align}
\tilde{P}_\phi^\mathrm{NG}(k)=\sum_{i=1}^nA_{\phi i}k_{*i}\delta(k-k_{*i})+f_\mathrm{NL}^2\sum_{i,j=1}^nA_{\phi i}A_{\phi j}\frac{k^2}{k_{*i}k_{*j}}\Theta_0(k_{*i}+k_{*j}-k)\Theta_0(k-|k_{*i}-k_{*j}|).
\end{align}
Now the product of $\tilde{P}_\phi^\mathrm{NG}(ku)\tilde{P}_\phi^\mathrm{NG}(kv)$ can be computed directly as
\begin{align}\label{eq:NGPs}
&\tilde{P}_\phi^\mathrm{NG}(ku)\tilde{P}_\phi^\mathrm{NG}(kv)=\sum_{i,l=1}^nA_{\phi i}A_{\phi l}k_{*i}k_{*l}\delta(ku-k_{*i})\delta(kv-k_{*l})\nonumber\\
&+f_\mathrm{NL}^2\sum_{i,l,m=1}^nA_{\phi i}A_{\phi l}A_{\phi m}k_{*i}\frac{k^2 v^2}{k_{*l}k_{*m}}\delta(ku-k_{*i})\Theta_0(k_{*l}+k_{*m}-kv)\Theta_0(kv-|k_{*l}-k_{*m}|)\nonumber\\
&+f_\mathrm{NL}^2\sum_{l,i,j=1}^nA_{\phi l}A_{\phi i}A_{\phi j}k_{*l}\frac{k^2 u^2}{k_{*i}k_{*j}}\delta(kv-k_{*l})\Theta_0(k_{*i}+k_{*j}-ku)\Theta_0(ku-|k_{*i}-k_{*j}|)\nonumber\\
&+f_\mathrm{NL}^4\sum_{i,j,l,m=1}^nA_{\phi i}A_{\phi j}A_{\phi l}A_{\phi m}\frac{k^2 u^2}{k_{*i}k_{*j}}\frac{k^2 v^2}{k_{*l}k_{*m}}\nonumber\\
&\times\Theta_0(k_{*i}+k_{*j}-ku)\Theta_0(ku-|k_{*i}-k_{*j}|)\Theta_0(k_{*l}+k_{*m}-kv)\Theta_0(kv-|k_{*l}-k_{*m}|).
\end{align}
To obtain the final integral of \eqref{eq:NGGW}, one can compute each term in \eqref{eq:NGPs}.
Then, the energy-density spectrum of induced GWs from non-Gaussian scalar perturbations with multiple $\delta$-peaks can be analytically obtained as
\begin{align}
\Omega_\mathrm{GW}^{n,\delta}(k)&=\frac{1}{24}\sum_{i,l=1}^nA_{\phi i}A_{\phi l}\frac{k^2}{k_{*i}k_{*l}}\mathcal{T}_{il}(k)\mathcal{F}^\delta(k;k_{*i},k_{*l})\nonumber\\
&+\frac{1}{24}f_\mathrm{NL}^2\sum_{i,l,m=1}^nA_{\phi i}A_{\phi l}A_{\phi m}\frac{k^3}{k_{*i}k_{*l}k_{*m}}\mathcal{T}_{ilm}(k)\mathcal{F}^\delta (k;k_{*i},k_{*l},k_{*m})\nonumber\\
&+\frac{1}{24}f_\mathrm{NL}^4\sum_{i,j,l,m=1}^nA_{\phi i}A_{\phi j}A_{\phi l}A_{\phi m}\frac{k^4}{k_{*i}k_{*j}k_{*l}k_{*m}}\mathcal{T}_{ijlm}(k)\mathcal{F}^\delta(k;k_{*i},k_{*j},k_{*l},k_{*m}).
\end{align}
Here we have introduced the following abbreviations
\begin{align}
\mathcal{T}_{il}(k)&\equiv \frac{k_{*i}^2 k_{*l}^2}{k^4} \mathcal{T}\left(\frac{k_{*i}}{k},\frac{k_{*l}}{k}\right);\\
\mathcal{T}_{ilm}(k)&\equiv 2\frac{k_{*i}^2}{k^2} \int_{\max\left(\frac{|k_{*l}-k_{*m}|}{k},\left|1-\frac{k_{*i}}{k}\right|\right)}^{\min\left(\frac{k_{*l}+k_{*m}}{k},1+\frac{k_{*i}}{k}\right)}\mathrm{d}v \  v^2 \mathcal{T}\left(\frac{k_{*i}}{k},v\right);\\
\mathcal{T}_{ijlm}(k)&\equiv\int_{\frac{|k_{*l}-k_{*m}|}{k}}^{\frac{k_{*l}+k_{*m}}{k}}\mathrm{d}v\int_{\max\left(\frac{|k_{*i}-k_{*j}|}{k},|1-v|\right)}^{\min\left(\frac{k_{*i}+k_{*j}}{k},1+v\right)}\mathrm{d}u \  u^2 v^2 \mathcal{T}(u,v),
\end{align}
and the definition of wave-vector factor
\begin{align}
\mathcal{F}^\delta(k;k_{*i},k_{*l})&\equiv\Theta_0(\max|\bm{k}_{*i}+\bm{k}_{*j}|-k)\Theta_0(k-\min|\bm{k}_{*i}+\bm{k}_{*j}|);\\
\mathcal{F}^\delta(k;k_{*i},k_{*l},k_{*m})&\equiv\Theta_0(\max|\bm{k}_{*i}+\bm{k}_{*l}+\bm{k}_{*m}|-k)\Theta_0(k-\min|\bm{k}_{*i}+\bm{k}_{*l}+\bm{k}_{*m}|);\\
\mathcal{F}^\delta(k;k_{*i},k_{*j},k_{*l},k_{*m})&\equiv\Theta_0(\max|\bm{k}_{*i}+\bm{k}_{*j}+\bm{k}_{*l}+\bm{k}_{*m}|-k)\Theta_0(k-\min|\bm{k}_{*i}+\bm{k}_{*j}+\bm{k}_{*l}+\bm{k}_{*m}|),
\end{align}
which come from the momentum conservation as we have discussed in the last section.

\section{Conclusions}\label{sec:con}

In this paper, the energy-density spectrum of induced GWs from a Gaussian scalar perturbations is studied analytically in details for two different type of peaks at small scales,
\begin{align}
\hbox{$\delta$-peak}:\quad&\tilde{P}_\phi(k)=\sum_{i=1}^nA_{\phi i}\delta(\ln\frac{k}{k_{*i}});\\
&\Omega_\mathrm{GW}^{n,\delta}(k)=\frac{1}{24}\sum_{i,j=1}^nA_{\phi i}A_{\phi j}\frac{k_{*i}k_{*j}}{k^2}\mathcal{T}\left(\frac{k_{*i}}{k},\frac{k_{*j}}{k}\right)\mathcal{F}^\delta(k;k_{*i},k_{*j});\\
\hbox{$\sigma$-peak}:\quad&\tilde{P}_{\phi}(k)=\sum\limits_{i=1}^n\frac{A_{\phi i}k_{*i}}{\sqrt{2\pi}\sigma}\mathrm{e}^{-\frac{(k-k_{*i})^2}{2\sigma^2}};\\
&\Omega_\mathrm{GW}^{n,\sigma}(k)=\frac{1}{24}\sum_{i,j=1}^nA_{\phi i}A_{\phi j}\frac{k_{*i}k_{*j}}{k^2}\mathcal{T}\left(\frac{k_{*i}}{k},\frac{k_{*j}}{k}\right)\mathcal{F}^\sigma(k;k_{*i},k_{*j}),
\end{align}
where $\mathcal{F^\delta}$ and $\mathcal{F}^\sigma$ are given by \eqref{eq:F_delta} and \eqref{eq:F_sigma}, respectively.
A multiple-peak structure in the energy-density spectrum of induced GWs is analytically identified at $k_{ij}=\frac{1}{\sqrt{3}}(k_{*i}+k_{*j})$, which can be interpreted as a consequence of resonant amplification.
Under the narrow-width approximation, the energy-density spectrum of induced GWs contains an universal factor that can be interpreted as the result of momentum conservation.
These observations also hold in the case of non-Gaussian scalar perturbations with multiple $\delta$-peaks, whose analytical expression of energy-density spectrum of induced GWs can be similarly obtained in a compact form.


\acknowledgments
SP and SJW want to thank the Institute of Theoretical Physics of CAS for the hospitality during their visit.
RGC is supported by the National Natural Science Foundation of China Grants Nos.11435006, 11647601, 11690022, 11821505, 11851302, and by the Strategic Priority Research Program of CAS Grant No.XDB23030100, and by the Key Research Program of Frontier Sciences of CAS\@.
SP is supported by the MEXT KAKENHI No.15H05888, and by the World Premier International Research Center Initiative (WPI Initiative), MEXT, Japan.
SJW is supported by the postdoctoral scholarship of Tufts University.

\appendix

\section{Triple $\delta$-peaks}

\begin{figure}[h]
	\includegraphics[width=0.3\textwidth]{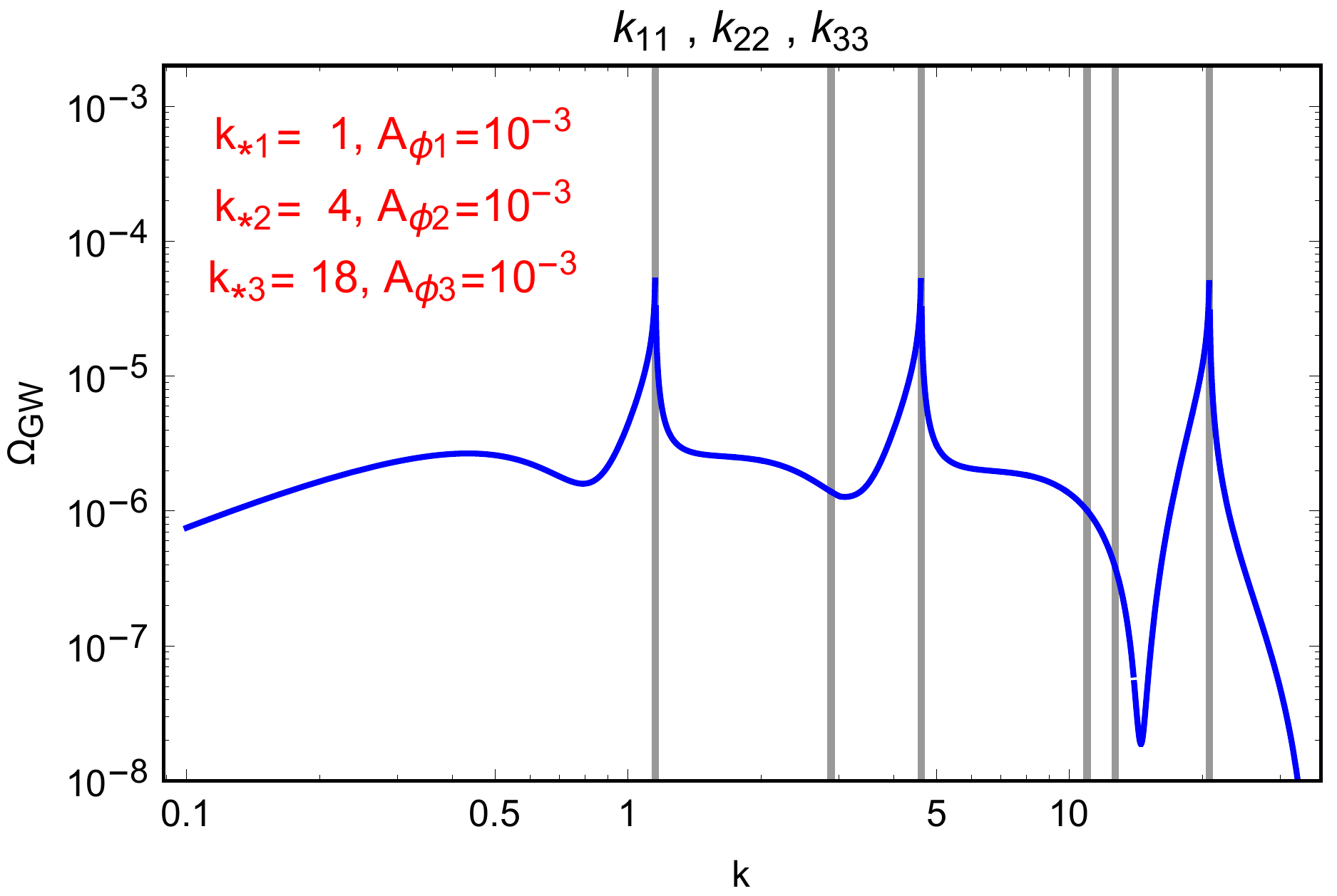}
	\includegraphics[width=0.3\textwidth]{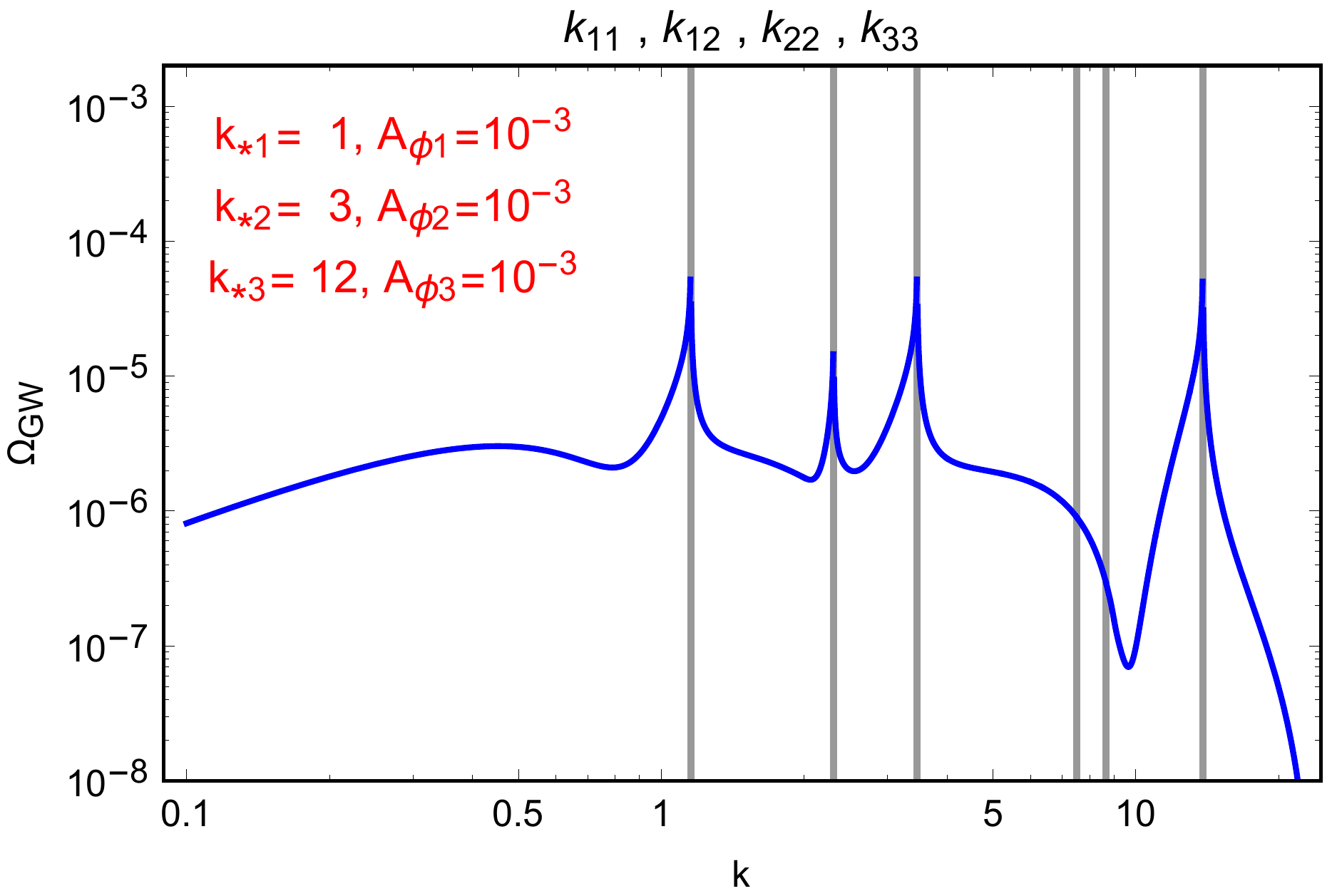}
	\includegraphics[width=0.3\textwidth]{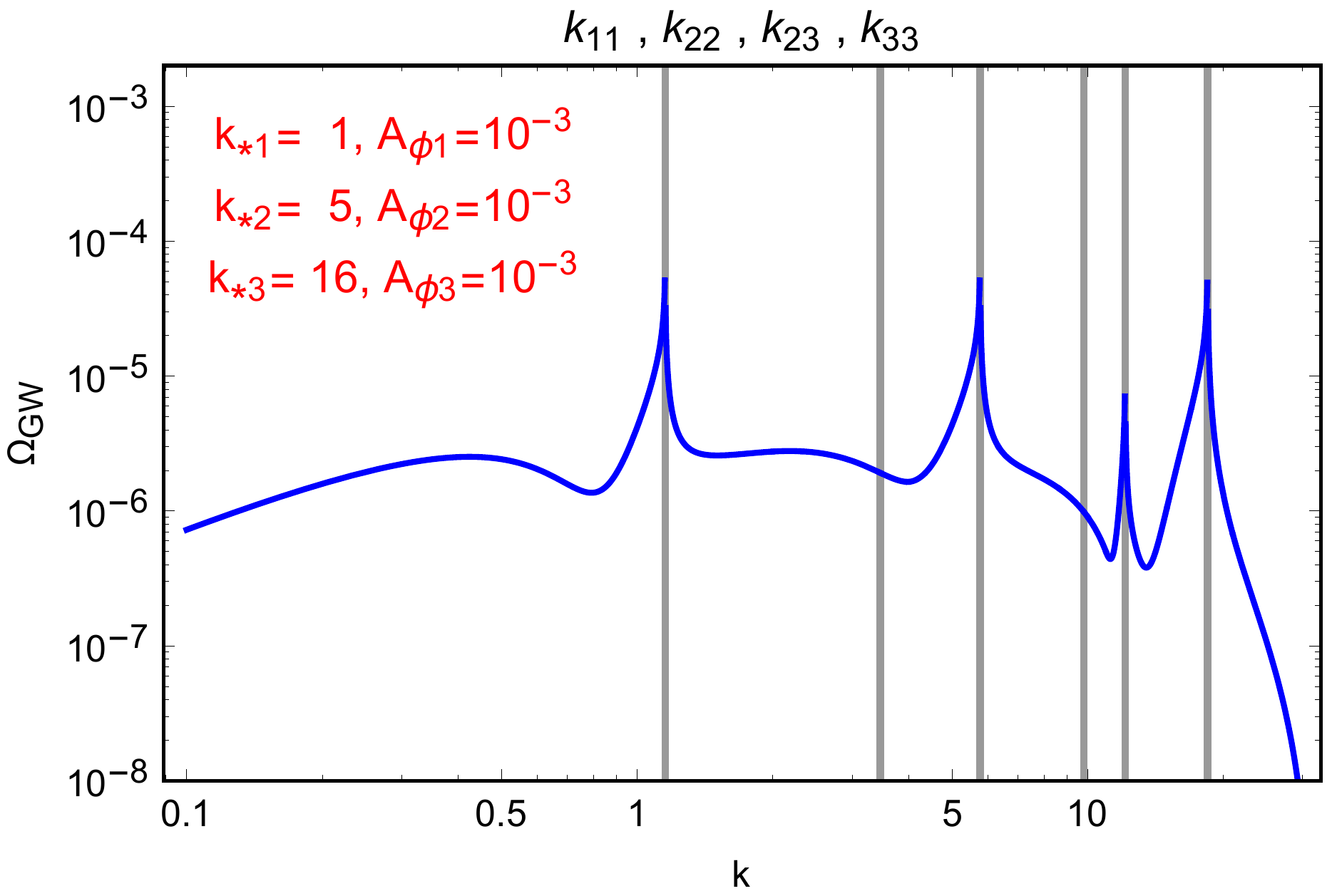}\\
	\includegraphics[width=0.3\textwidth]{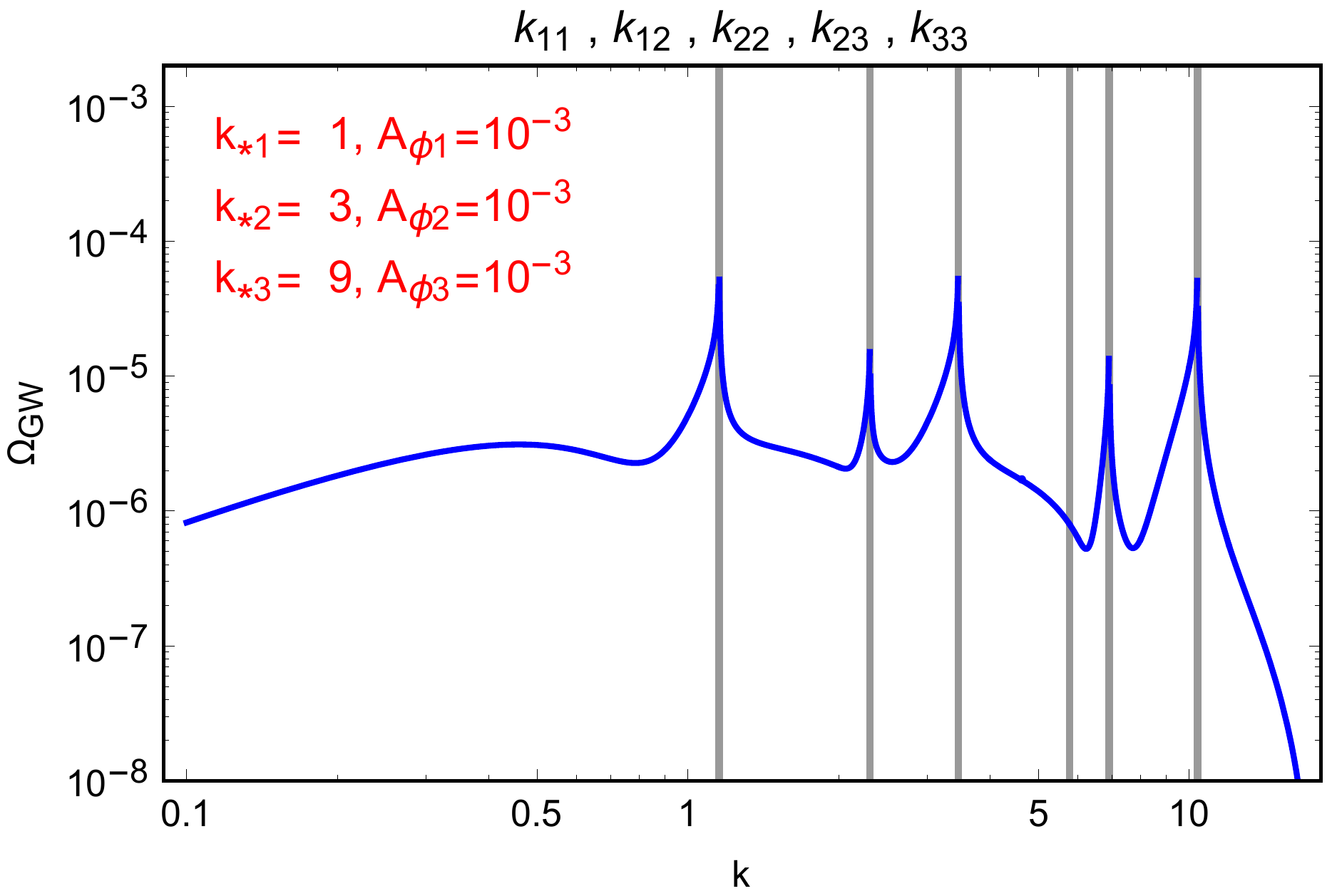}
	\includegraphics[width=0.3\textwidth]{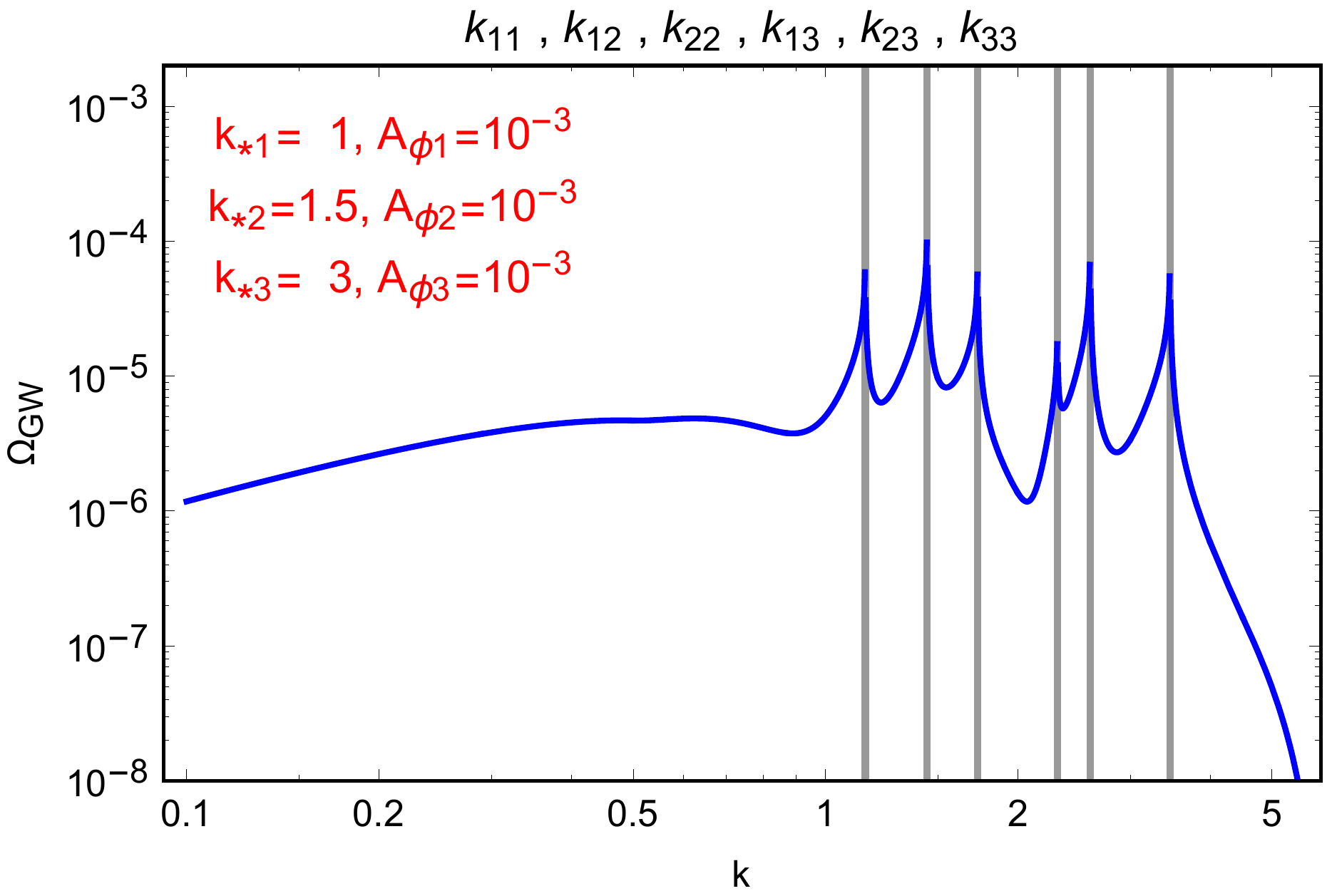}
	\includegraphics[width=0.3\textwidth]{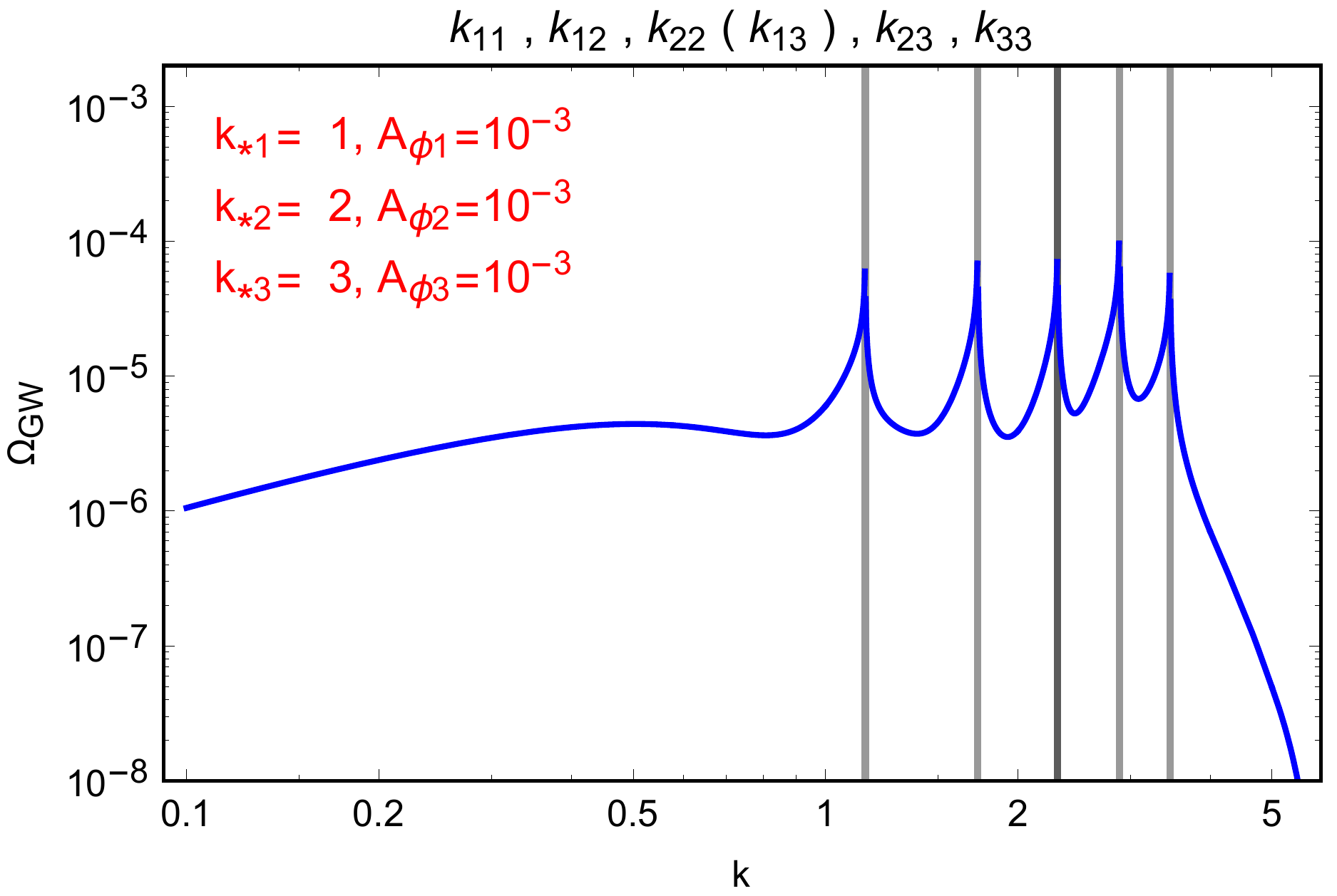}\\
	\caption{The energy-density spectrum of induced GWs from scalar perturbations with triple $\delta$-peaks at $k_{*1}<k_{*2}<k_{*3}$.
The gray lines denote the positions of those would-be peaks at $k_{ij}$ with $i,j=1,2,3$.}\label{fig:3delta}
\end{figure}

We present in Fig.\ref{fig:3delta} all the possible cases of induced GWs from Gaussian scalar perturbations with triple $\delta$-peaks at $k_{*1}<k_{*2}<k_{*3}$, where the gray lines denote the positions of those would-be peaks at
\begin{align}
k_{ij}=\frac{k_{*i}+k_{*j}}{\sqrt{3}},\quad i,j=1,2,3.
\end{align}
The purpose of this appendix is to show that, there are at most $C_{n+1}^2=C_4^2=6$ and at least $n=3$ peaks in the energy-density spectrum.
In the first panel, the peaks at position of $k_{12}$ , $k_{13}$ and $k_{23}$ are vanish because they violate the momentum conservation condition $|k_{*i}-k_{*j}|<k_{ij}<k_{*i}+k_{*j}$, similar cases also occur in the other panels for the vanishing peaks.
In the last panel, $k_{22}=k_{13}$, which makes these two peaks overlap.

\section{Constraints from PBH and GW}

\begin{figure}
\centering
\includegraphics[width=\textwidth]{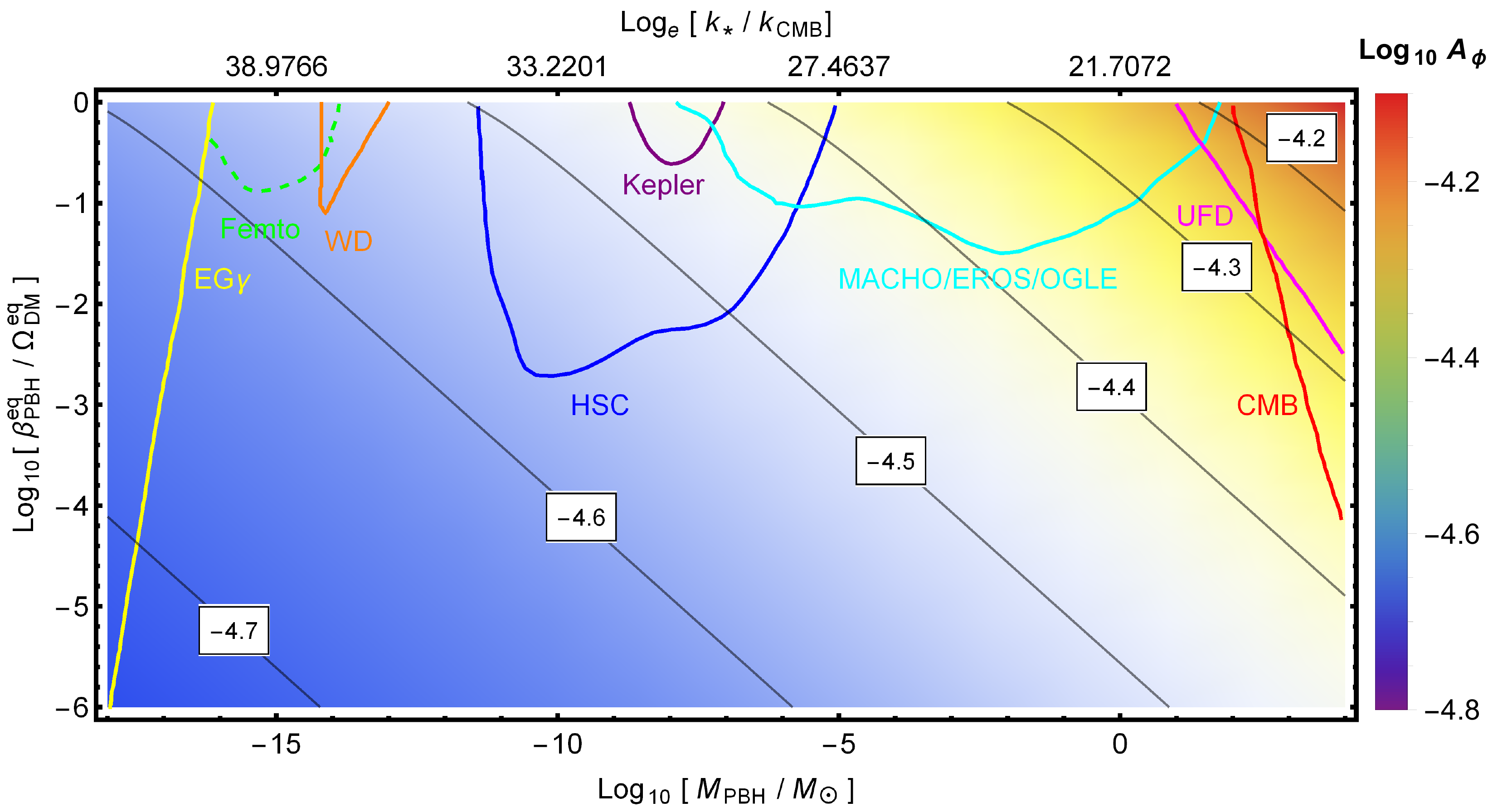}\\
\includegraphics[width=0.49\textwidth]{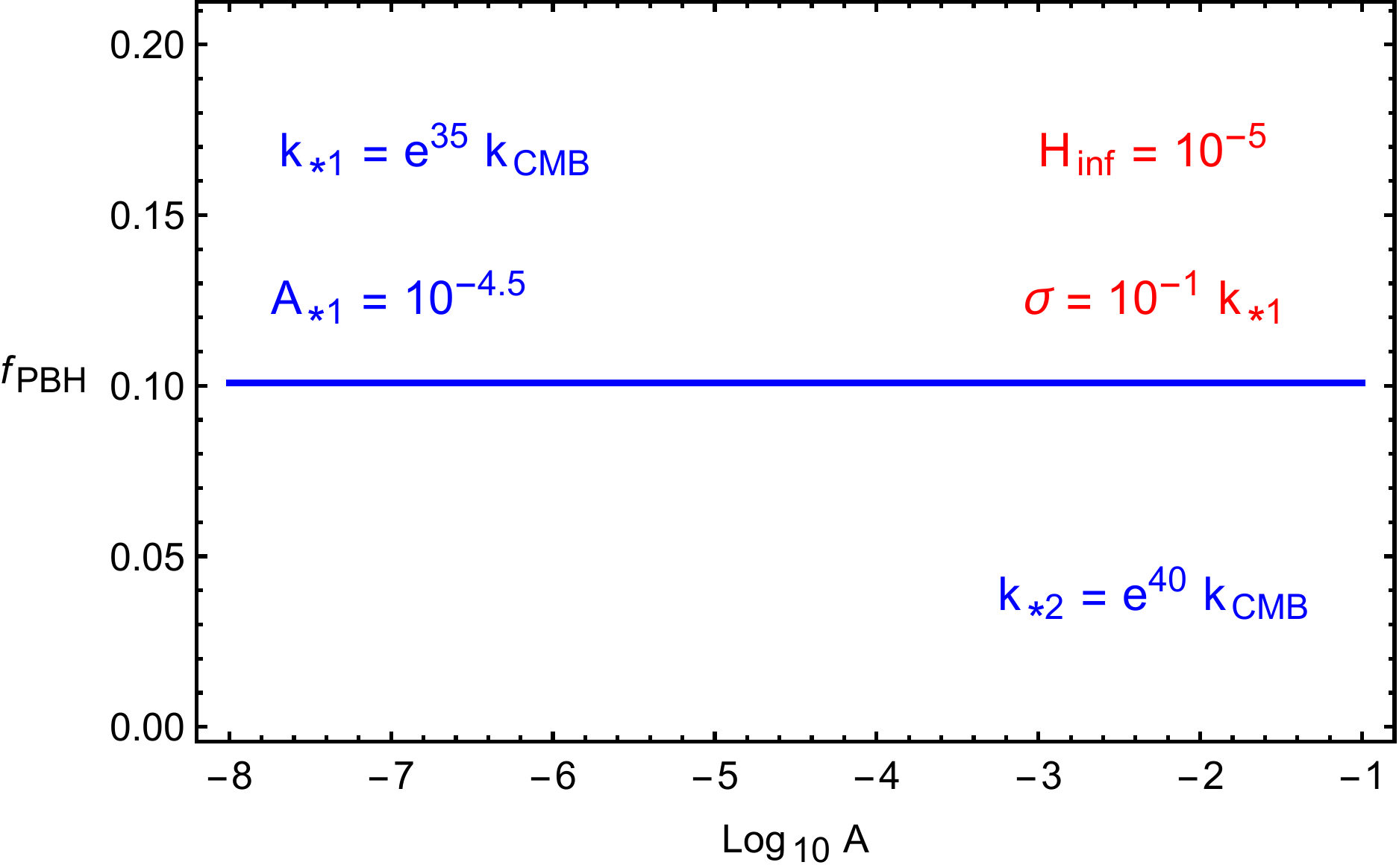}
\includegraphics[width=0.49\textwidth]{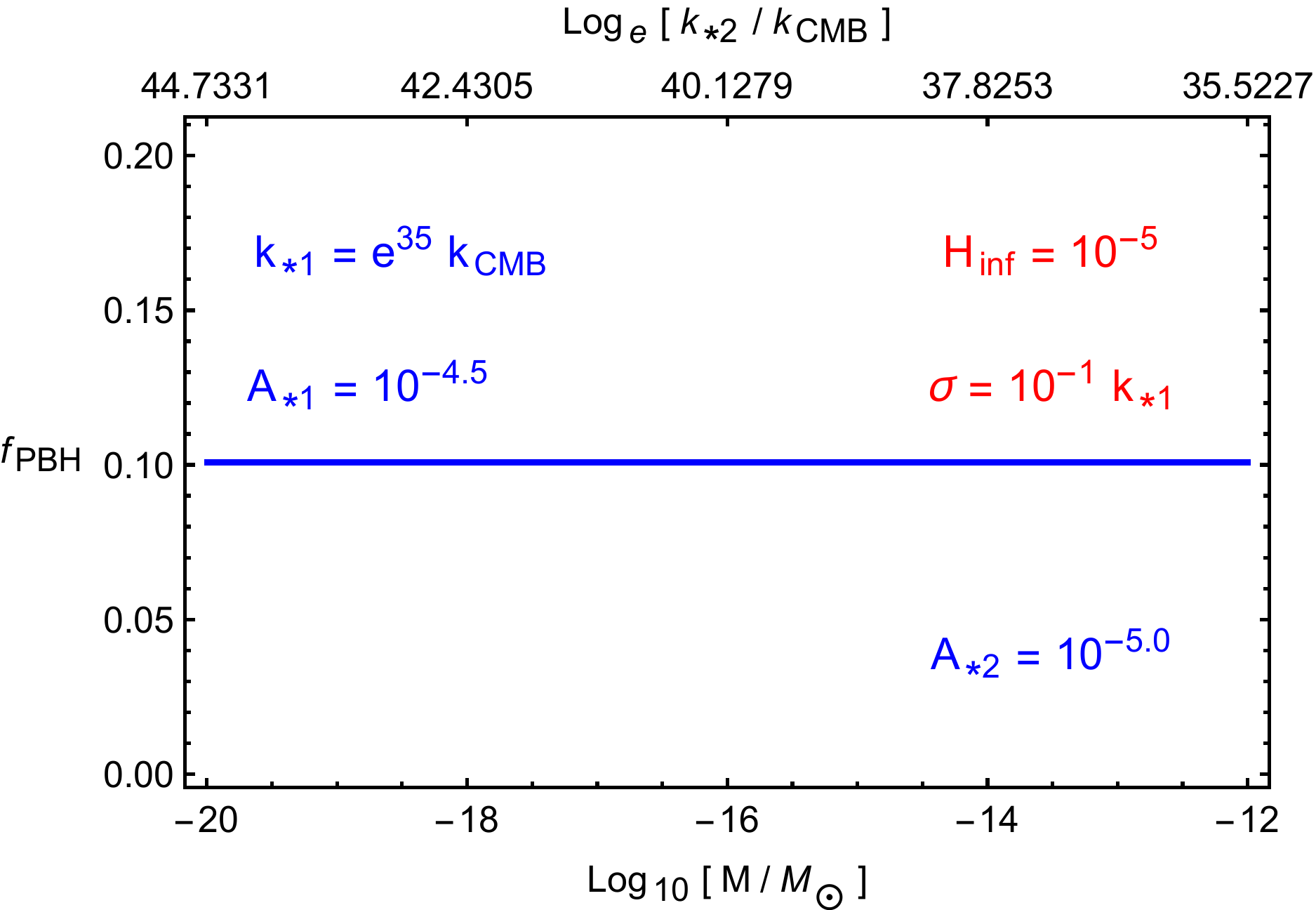}\\
\caption{\textit{Top}: Constraints on the peak position $k_*$ and corresponding amplitude $A_\phi$ of scalar perturbation that leads to the formation of PBH\@.
We have adopted the PBH constraints from the extragalactic photon (EG$\gamma$\cite{Carr:2009jm}), femtolensing of gamma-ray burst (Femto\cite{Barnacka:2012bm}, however, see\cite{Katz:2018zrn} for criticism), white dwarf explosions (WD\cite{Graham:2015apa}), microlensing from Subaru Hyper Suprime-Cam (HSC\cite{Niikura:2017zjd}), MACHO\cite{Allsman:2000kg}, EROS\cite{Tisserand:2006zx} and OGLE\cite{Wyrzykowski:2011tr}, ultrafaint dwarfs (UFD\cite{Brandt:2016aco}) and CMB\cite{Ali-Haimoud:2016mbv}.
\textit{Bottom}: The PBH abundance in the presence of a second peak in the scalar perturbations, where the position and amplitude of the first peak have been fixed for clarity.
The position and amplitude of the second peak have been fixed separately in the bottom left and right panels, respectively.
The PBH abundance is insensitive to the second peak in scalar perturbation.}\label{fig:PBH}
\end{figure}

So far the peak position and corresponding amplitude of scalar perturbation are left as free parameters, which could be constrained by the current non-detection of PBH\@.
For scalar perturbation with a single $\sigma$-peak at $k_*$ of width $\sigma=10^{-1}k_*$, the required peak amplitude $A_\phi$ for the formation of PBH can be constrained in the mass-fraction plane as shown in the top panel of Fig.\ref{fig:PBH}.
The details and conventions for this calculation can be found in, e.g.\cite{Cai:2018rqf}(see also\cite{Wang:2019kaf}), where we have assumed a constant inflationary scale $H_\mathrm{inf}=10^{-5} M_\mathrm{Pl}$, an illustrative e-folding number $N_\mathrm{CMB}=60$, an instantaneous reheating history $N_\mathrm{reh}=0$ and a PBH formation threshold $\delta_c=0.086$\cite{Harada:2013epa}.
As one can see, there are currently two windows for PBH making up all DM with following choices for the parameters:
\begin{align}
10^{-16}\,M_\odot\lesssim M_\mathrm{PBH}\lesssim10^{-14}\,M_\odot :&\quad\mathrm{e}^{37.83}\lesssim\frac{k_*}{k_\mathrm{CMB}}\lesssim\mathrm{e}^{40.13},\quad 10^{-4.54}\gtrsim A_\phi\gtrsim 10^{-4.57};\label{eq:window1}\\
10^{-13}\,M_\odot\lesssim M_\mathrm{PBH}\lesssim10^{-11}\,M_\odot :&\quad\mathrm{e}^{34.37}\lesssim\frac{k_*}{k_\mathrm{CMB}}\lesssim\mathrm{e}^{36.67},\quad 10^{-4.49}\gtrsim A_\phi\gtrsim 10^{-4.52}.\label{eq:window2}
\end{align}

One can also show that the PBH abundance is insensitive to the second peak in the scalar perturbations.
To see this, we first fix the position and amplitude of the first peak in the scalar perturbations at $k_{*1}=\mathrm{e}^{35}k_\mathrm{CMB}, A_1=10^{-4.5}$.
Then one can change separately the position and amplitude of the second peak in the scalar perturbations, respectively.
The PBH abundance is unchanged as shown in the bottom panels of Fig.\ref{fig:PBH}.
The reason for this insensitivity lies in the configuration of peak width $\sigma=\epsilon k_{*1}$ for all peaks, then the relative width of second peak $k_{*2}$ is more narrow than the first peak $k_{*1}$, namely $\sigma/k_{*2}=\epsilon k_{*1}/k_{*2}<\epsilon=\sigma/k_{*1}$, therefore, the PBH abundance (the area below the curve in the PBH constraint plane) is primarily determined by the first peak in the scalar perturbations.
However, the second peak could play a more important role when the width of each peak could be configured separately, which will be explored in future.

\begin{figure}
\centering
\includegraphics[width=0.49\textwidth]{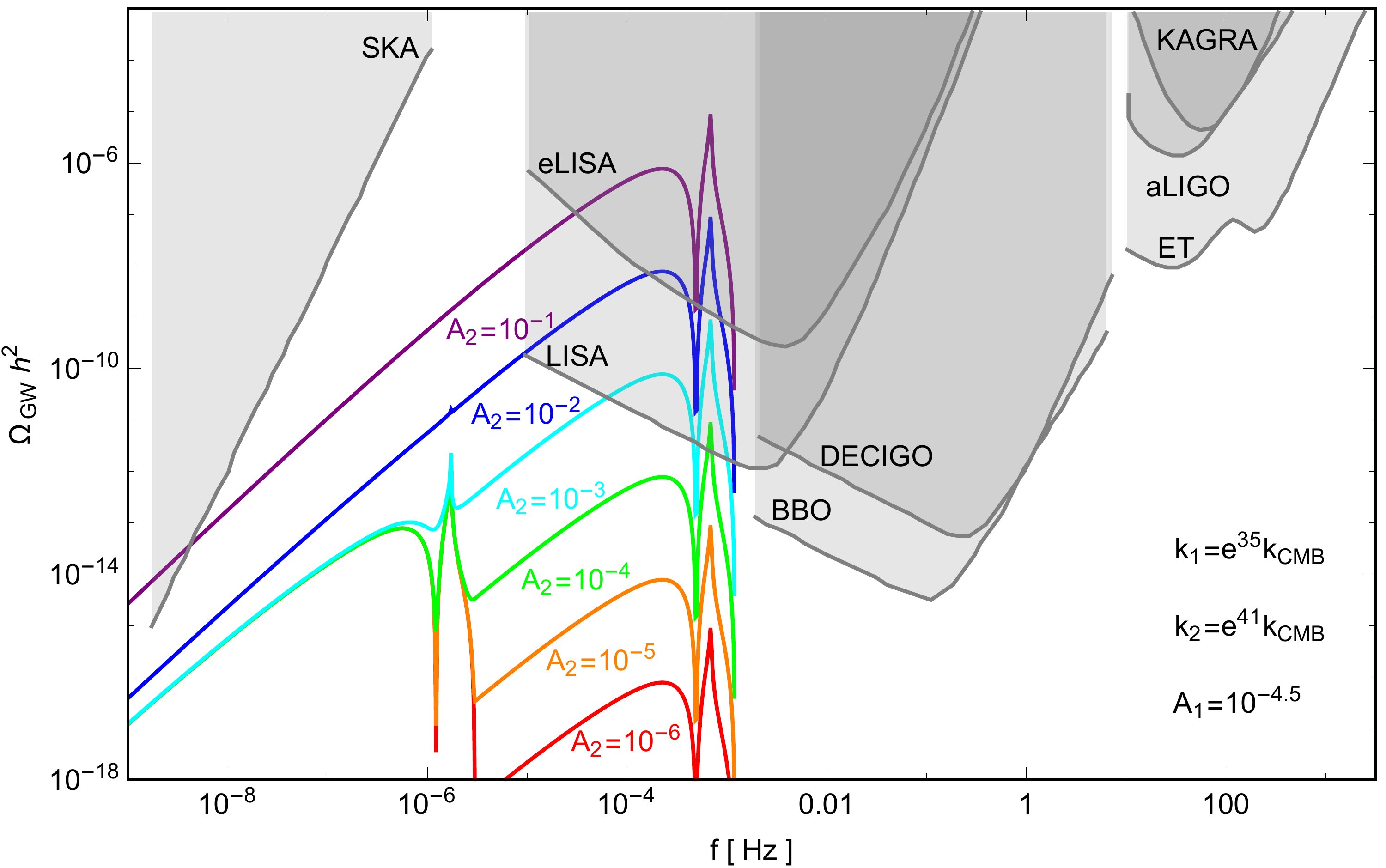}
\includegraphics[width=0.49\textwidth]{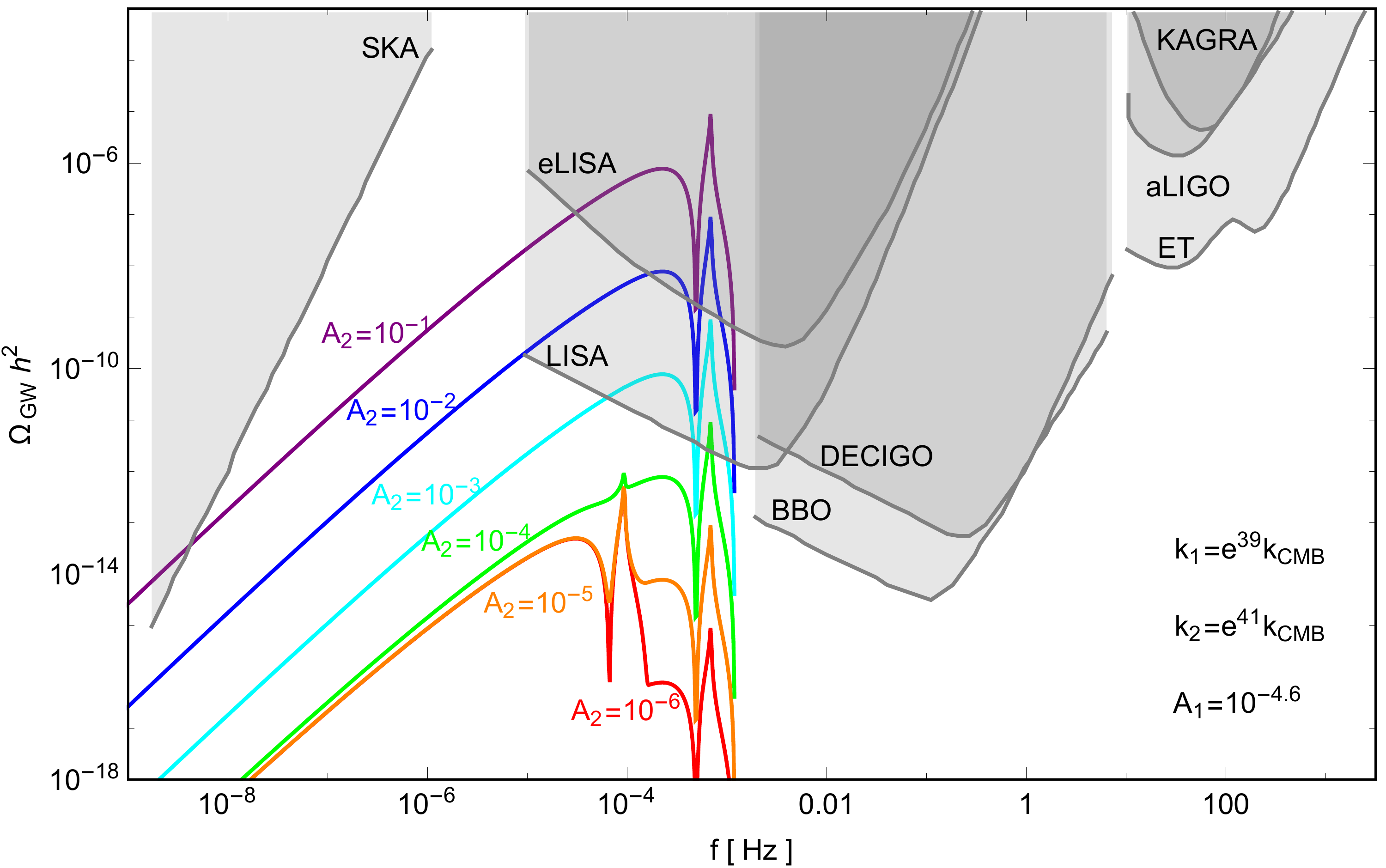}\\
\includegraphics[width=0.49\textwidth]{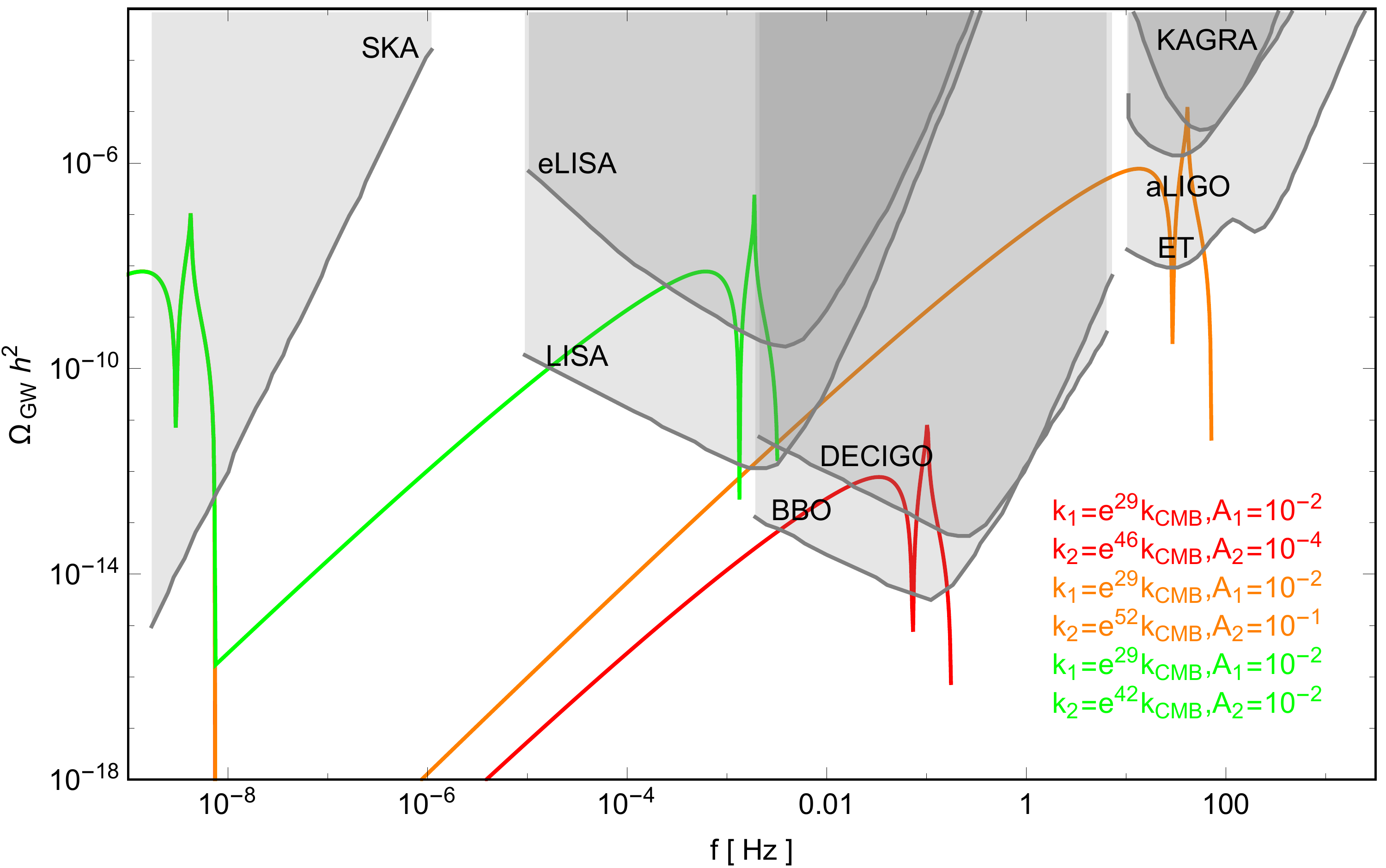}
\includegraphics[width=0.49\textwidth]{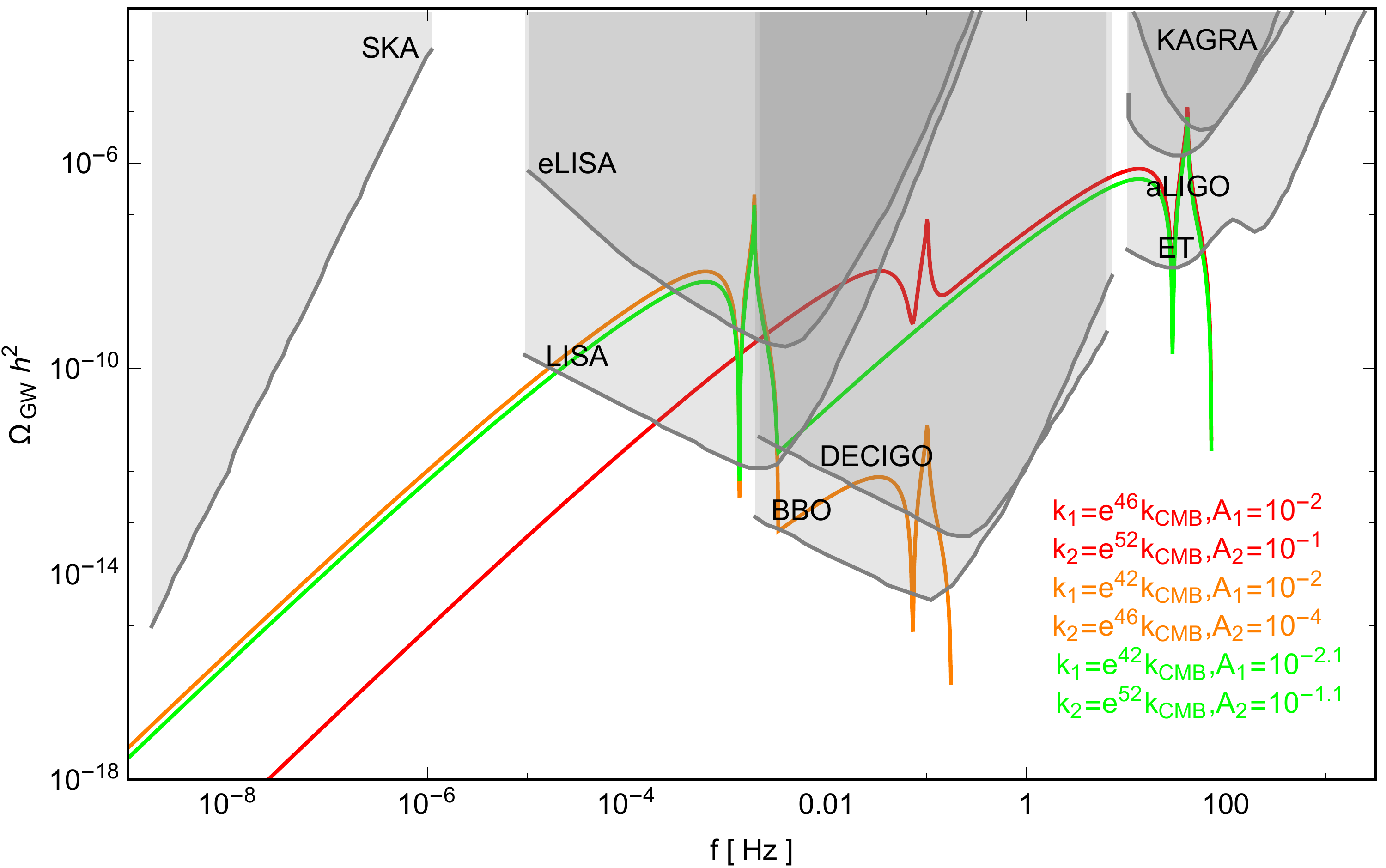}\\
\caption{In the first two panels, we present the observed induced GWs from scalar perturbation with double $\delta$-peaks that could later collapse into PBHs constrained by \ref{fig:PBH}.
The first panel is subjected to the mass window \eqref{eq:window2}, and the second panel is subjected to the mass window \eqref{eq:window1}.
The amplitude of the second peak is varied from $10^{-6}$ to $10^{-1}$.
In the last two panels, we forget about the PBH constraint, therefore,  we are free to adjust the position and amplitude of both peaks so that the induced GWs could probe the sensitivity ranges of each GW detector.}\label{fig:GW}
\end{figure}
Nevertheless, the effect of second peak in the scalar perturbation could manifest itself on the induced GWs, which is illustrated in the Fig.\ref{fig:GW} for a scalar perturbation with double $\delta$-peaks.
In the first two panels, the first peak of scalar perturbation is fixed to meet the PBH constraint in Fig.\ref{fig:PBH}, where the first panel is subjected to the mass window $10^{-13}\,M_\odot\lesssim M_\mathrm{PBH}\lesssim10^{-11}\,M_\odot$, and the second panel is subjected to the mass window $10^{-16}\,M_\odot\lesssim M_\mathrm{PBH}\lesssim10^{-14}\,M_\odot$.
With increasing amplitude of the second peak in scalar perturbation, the first peak in the induced GWs becomes invisible.
In the last two panels that are not subjected to any PBH constraint, both the position and amplitude of each peak of scalar perturbation are free to choose so that the resulting induced GWs could overlap with each sensitivity range of all GW detectors.

\bibliographystyle{JHEP}
\bibliography{ref}

\providecommand{\href}[2]{#2}\begingroup\raggedright\begin{thebibliography}{10}

\bibitem{Abbott:2016blz}
{\scshape Virgo, LIGO Scientific} collaboration, B.~P. Abbott et~al.,
  \emph{{Observation of Gravitational Waves from a Binary Black Hole Merger}},
  \href{http://dx.doi.org/10.1103/PhysRevLett.116.061102}{\emph{Phys. Rev.
  Lett.} {\bf 116} (2016) 061102},
  [\href{https://arxiv.org/abs/1602.03837}{{\tt 1602.03837}}].

\bibitem{TheLIGOScientific:2017qsa}
{\scshape Virgo, LIGO Scientific} collaboration, B.~Abbott et~al.,
  \emph{{GW170817: Observation of Gravitational Waves from a Binary Neutron
  Star Inspiral}},
  \href{http://dx.doi.org/10.1103/PhysRevLett.119.161101}{\emph{Phys. Rev.
  Lett.} {\bf 119} (2017) 161101},
  [\href{https://arxiv.org/abs/1710.05832}{{\tt 1710.05832}}].

\bibitem{Caprini:2018mtu}
C.~Caprini and D.~G. Figueroa, \emph{{Cosmological Backgrounds of Gravitational
  Waves}}, \href{http://dx.doi.org/10.1088/1361-6382/aac608}{\emph{Class.
  Quant. Grav.} {\bf 35} (2018) 163001},
  [\href{https://arxiv.org/abs/1801.04268}{{\tt 1801.04268}}].

\bibitem{Cai:2017cbj}
R.-G. Cai, Z.~Cao, Z.-K. Guo, S.-J. Wang and T.~Yang, \emph{{The
  Gravitational-Wave Physics}},
  \href{http://dx.doi.org/10.1093/nsr/nwx029}{\emph{National Science Review}
  {\bf 4} (2017) 687--706}, [\href{https://arxiv.org/abs/1703.00187}{{\tt
  1703.00187}}].

\bibitem{Guzzetti:2016mkm}
C.~Guzzetti, M., N.~Bartolo, M.~Liguori and S.~Matarrese, \emph{{Gravitational
  waves from inflation}},
  \href{http://dx.doi.org/10.1393/ncr/i2016-10127-1}{\emph{Riv. Nuovo Cim.}
  {\bf 39} (2016) 399--495}, [\href{https://arxiv.org/abs/1605.01615}{{\tt
  1605.01615}}].

\bibitem{Bartolo:2016ami}
N.~Bartolo et~al., \emph{{Science with the space-based interferometer LISA. IV:
  Probing inflation with gravitational waves}},
  \href{http://dx.doi.org/10.1088/1475-7516/2016/12/026}{\emph{JCAP} {\bf 1612}
  (2016) 026}, [\href{https://arxiv.org/abs/1610.06481}{{\tt 1610.06481}}].

\bibitem{Boyle:2003km}
L.~A. Boyle, P.~J. Steinhardt and N.~Turok, \emph{{The Cosmic gravitational
  wave background in a cyclic universe}},
  \href{http://dx.doi.org/10.1103/PhysRevD.69.127302}{\emph{Phys. Rev.} {\bf
  D69} (2004) 127302}, [\href{https://arxiv.org/abs/hep-th/0307170}{{\tt
  hep-th/0307170}}].

\bibitem{Gasperini:2002bn}
M.~Gasperini and G.~Veneziano, \emph{{The Pre - big bang scenario in string
  cosmology}},
  \href{http://dx.doi.org/10.1016/S0370-1573(02)00389-7}{\emph{Phys. Rept.}
  {\bf 373} (2003) 1--212}, [\href{https://arxiv.org/abs/hep-th/0207130}{{\tt
  hep-th/0207130}}].

\bibitem{Lehners:2008vx}
J.-L. Lehners, \emph{{Ekpyrotic and Cyclic Cosmology}},
  \href{http://dx.doi.org/10.1016/j.physrep.2008.06.001}{\emph{Phys. Rept.}
  {\bf 465} (2008) 223--263}, [\href{https://arxiv.org/abs/0806.1245}{{\tt
  0806.1245}}].

\bibitem{Battefeld:2014uga}
D.~Battefeld and P.~Peter, \emph{{A Critical Review of Classical Bouncing
  Cosmologies}},
  \href{http://dx.doi.org/10.1016/j.physrep.2014.12.004}{\emph{Phys. Rept.}
  {\bf 571} (2015) 1--66}, [\href{https://arxiv.org/abs/1406.2790}{{\tt
  1406.2790}}].

\bibitem{Cai:2014bea}
Y.-F. Cai, \emph{{Exploring Bouncing Cosmologies with Cosmological Surveys}},
  \href{http://dx.doi.org/10.1007/s11433-014-5512-3}{\emph{Sci. China Phys.
  Mech. Astron.} {\bf 57} (2014) 1414--1430},
  [\href{https://arxiv.org/abs/1405.1369}{{\tt 1405.1369}}].

\bibitem{Brandenberger:2016vhg}
R.~Brandenberger and P.~Peter, \emph{{Bouncing Cosmologies: Progress and
  Problems}}, \href{http://dx.doi.org/10.1007/s10701-016-0057-0}{\emph{Found.
  Phys.} {\bf 47} (2017) 797--850},
  [\href{https://arxiv.org/abs/1603.05834}{{\tt 1603.05834}}].

\bibitem{Allahverdi:2010xz}
R.~Allahverdi, R.~Brandenberger, F.-Y. Cyr-Racine and A.~Mazumdar,
  \emph{{Reheating in Inflationary Cosmology: Theory and Applications}},
  \href{http://dx.doi.org/10.1146/annurev.nucl.012809.104511}{\emph{Ann. Rev.
  Nucl. Part. Sci.} {\bf 60} (2010) 27--51},
  [\href{https://arxiv.org/abs/1001.2600}{{\tt 1001.2600}}].

\bibitem{Amin:2014eta}
M.~A. Amin, M.~P. Hertzberg, D.~I. Kaiser and J.~Karouby,
  \emph{{Nonperturbative Dynamics Of Reheating After Inflation: A Review}},
  \href{http://dx.doi.org/10.1142/S0218271815300037}{\emph{Int. J. Mod. Phys.}
  {\bf D24} (2014) 1530003}, [\href{https://arxiv.org/abs/1410.3808}{{\tt
  1410.3808}}].

\bibitem{Binetruy:2012ze}
P.~Binetruy, A.~Bohe, C.~Caprini and J.-F. Dufaux, \emph{{Cosmological
  Backgrounds of Gravitational Waves and eLISA/NGO: Phase Transitions, Cosmic
  Strings and Other Sources}},
  \href{http://dx.doi.org/10.1088/1475-7516/2012/06/027}{\emph{JCAP} {\bf 1206}
  (2012) 027}, [\href{https://arxiv.org/abs/1201.0983}{{\tt 1201.0983}}].

\bibitem{Caprini:2015zlo}
C.~Caprini et~al., \emph{{Science with the space-based interferometer eLISA.
  II: Gravitational waves from cosmological phase transitions}},
  \href{http://dx.doi.org/10.1088/1475-7516/2016/04/001}{\emph{JCAP} {\bf 1604}
  (2016) 001}, [\href{https://arxiv.org/abs/1512.06239}{{\tt 1512.06239}}].

\bibitem{Ananda:2006af}
K.~N. Ananda, C.~Clarkson and D.~Wands, \emph{{The Cosmological gravitational
  wave background from primordial density perturbations}},
  \href{http://dx.doi.org/10.1103/PhysRevD.75.123518}{\emph{Phys. Rev.} {\bf
  D75} (2007) 123518}, [\href{https://arxiv.org/abs/gr-qc/0612013}{{\tt
  gr-qc/0612013}}].

\bibitem{Baumann:2007zm}
D.~Baumann, P.~J. Steinhardt, K.~Takahashi and K.~Ichiki, \emph{{Gravitational
  Wave Spectrum Induced by Primordial Scalar Perturbations}},
  \href{http://dx.doi.org/10.1103/PhysRevD.76.084019}{\emph{Phys. Rev.} {\bf
  D76} (2007) 084019}, [\href{https://arxiv.org/abs/hep-th/0703290}{{\tt
  hep-th/0703290}}].

\bibitem{Saito:2008jc}
R.~Saito and J.~Yokoyama, \emph{{Gravitational wave background as a probe of
  the primordial black hole abundance}},
  \href{http://dx.doi.org/10.1103/PhysRevLett.102.161101,
  10.1103/PhysRevLett.107.069901}{\emph{Phys. Rev. Lett.} {\bf 102} (2009)
  161101}, [\href{https://arxiv.org/abs/0812.4339}{{\tt 0812.4339}}].

\bibitem{Saito:2009jt}
R.~Saito and J.~Yokoyama, \emph{{Gravitational-Wave Constraints on the
  Abundance of Primordial Black Holes}},
  \href{http://dx.doi.org/10.1143/PTP.126.351, 10.1143/PTP.123.867}{\emph{Prog.
  Theor. Phys.} {\bf 123} (2010) 867--886},
  [\href{https://arxiv.org/abs/0912.5317}{{\tt 0912.5317}}].

\bibitem{Garcia-Bellido:2017aan}
J.~Garcia-Bellido, M.~Peloso and C.~Unal, \emph{{Gravitational Wave signatures
  of inflationary models from Primordial Black Hole Dark Matter}},
  \href{http://dx.doi.org/10.1088/1475-7516/2017/09/013}{\emph{JCAP} {\bf 1709}
  (2017) 013}, [\href{https://arxiv.org/abs/1707.02441}{{\tt 1707.02441}}].

\bibitem{Ando:2017veq}
K.~Ando, K.~Inomata, M.~Kawasaki, K.~Mukaida and T.~T. Yanagida,
  \emph{{Primordial black holes for the LIGO events in the axionlike curvaton
  model}}, \href{http://dx.doi.org/10.1103/PhysRevD.97.123512}{\emph{Phys.
  Rev.} {\bf D97} (2018) 123512}, [\href{https://arxiv.org/abs/1711.08956}{{\tt
  1711.08956}}].

\bibitem{Espinosa:2018eve}
J.~R. Espinosa, D.~Racco and A.~Riotto, \emph{{A Cosmological Signature of the
  SM Higgs Instability: Gravitational Waves}},
  \href{http://dx.doi.org/10.1088/1475-7516/2018/09/012}{\emph{JCAP} {\bf 1809}
  (2018) 012}, [\href{https://arxiv.org/abs/1804.07732}{{\tt 1804.07732}}].

\bibitem{Kohri:2018awv}
K.~Kohri and T.~Terada, \emph{{Semianalytic calculation of gravitational wave
  spectrum nonlinearly induced from primordial curvature perturbations}},
  \href{http://dx.doi.org/10.1103/PhysRevD.97.123532}{\emph{Phys. Rev.} {\bf
  D97} (2018) 123532}, [\href{https://arxiv.org/abs/1804.08577}{{\tt
  1804.08577}}].

\bibitem{Cai:2018dig}
R.-g. Cai, S.~Pi and M.~Sasaki, \emph{{Gravitational Waves Induced by
  non-Gaussian Scalar Perturbations}},
  \href{https://arxiv.org/abs/1810.11000}{{\tt 1810.11000}}.

\bibitem{Bartolo:2018evs}
N.~Bartolo, V.~De~Luca, G.~Franciolini, A.~Lewis, M.~Peloso and A.~Riotto,
  \emph{{The Primordial Black Hole Dark Matter - LISA Serendipity}},
  \href{https://arxiv.org/abs/1810.12218}{{\tt 1810.12218}}.

\bibitem{Bartolo:2018rku}
N.~Bartolo, V.~De~Luca, G.~Franciolini, M.~Peloso, D.~Racco and A.~Riotto,
  \emph{{Testing Primordial Black Holes as Dark Matter through LISA}},
  \href{https://arxiv.org/abs/1810.12224}{{\tt 1810.12224}}.

\bibitem{Unal:2018yaa}
C.~Unal, \emph{{Imprints of Primordial Non-Gaussianity on Gravitational Wave
  Spectrum}}, \href{http://dx.doi.org/10.1103/PhysRevD.99.041301}{\emph{Phys.
  Rev.} {\bf D99} (2019) 041301}, [\href{https://arxiv.org/abs/1811.09151}{{\tt
  1811.09151}}].

\bibitem{Byrnes:2018txb}
C.~T. Byrnes, P.~S. Cole and S.~P. Patil, \emph{{Steepest growth of the power
  spectrum and primordial black holes}},
  \href{https://arxiv.org/abs/1811.11158}{{\tt 1811.11158}}.

\bibitem{Inomata:2018epa}
K.~Inomata and T.~Nakama, \emph{{Gravitational waves induced by scalar
  perturbations as probes of the small-scale primordial spectrum}},
  \href{https://arxiv.org/abs/1812.00674}{{\tt 1812.00674}}.

\bibitem{Akrami:2018odb}
{\scshape Planck} collaboration, Y.~Akrami et~al., \emph{{Planck 2018 results.
  X. Constraints on inflation}},  \href{https://arxiv.org/abs/1807.06211}{{\tt
  1807.06211}}.

\bibitem{Ade:2018gkx}
{\scshape BICEP2, Keck Array} collaboration, P.~A.~R. Ade et~al., \emph{{BICEP2
  / Keck Array x: Constraints on Primordial Gravitational Waves using Planck,
  WMAP, and New BICEP2/Keck Observations through the 2015 Season}},
  \href{http://dx.doi.org/10.1103/PhysRevLett.121.221301}{\emph{Phys. Rev.
  Lett.} {\bf 121} (2018) 221301},
  [\href{https://arxiv.org/abs/1810.05216}{{\tt 1810.05216}}].

\bibitem{Matsumura:2013aja}
T.~Matsumura et~al., \emph{{Mission design of LiteBIRD}},
  \href{https://arxiv.org/abs/1311.2847}{{\tt 1311.2847}}.

\bibitem{Li:2017drr}
H.~Li et~al., \emph{{Probing Primordial Gravitational Waves: Ali CMB
  Polarization Telescope}},  \href{https://arxiv.org/abs/1710.03047}{{\tt
  1710.03047}}.

\bibitem{Crowder:2005nr}
J.~Crowder and N.~J. Cornish, \emph{{Beyond LISA: Exploring future
  gravitational wave missions}},
  \href{http://dx.doi.org/10.1103/PhysRevD.72.083005}{\emph{Phys. Rev.} {\bf
  D72} (2005) 083005}, [\href{https://arxiv.org/abs/gr-qc/0506015}{{\tt
  gr-qc/0506015}}].

\bibitem{Corbin:2005ny}
V.~Corbin and N.~J. Cornish, \emph{{Detecting the cosmic gravitational wave
  background with the big bang observer}},
  \href{http://dx.doi.org/10.1088/0264-9381/23/7/014}{\emph{Class. Quant.
  Grav.} {\bf 23} (2006) 2435--2446},
  [\href{https://arxiv.org/abs/gr-qc/0512039}{{\tt gr-qc/0512039}}].

\bibitem{Kawamura:2006up}
S.~Kawamura et~al., \emph{{The Japanese space gravitational wave antenna
  DECIGO}}, \href{http://dx.doi.org/10.1088/0264-9381/23/8/S17}{\emph{Class.
  Quant. Grav.} {\bf 23} (2006) S125--S132}.

\bibitem{Kawamura:2011zz}
S.~Kawamura et~al., \emph{{The Japanese space gravitational wave antenna:
  DECIGO}}, \href{http://dx.doi.org/10.1088/0264-9381/28/9/094011}{\emph{Class.
  Quant. Grav.} {\bf 28} (2011) 094011}.

\bibitem{AmaroSeoane:2012km}
P.~Amaro-Seoane et~al., \emph{{eLISA/NGO: Astrophysics and cosmology in the
  gravitational-wave millihertz regime}}, {\emph{GW Notes} {\bf 6} (2013)
  4--110}, [\href{https://arxiv.org/abs/1201.3621}{{\tt 1201.3621}}].

\bibitem{AmaroSeoane:2012je}
P.~Amaro-Seoane et~al., \emph{{Low-frequency gravitational-wave science with
  eLISA/NGO}},
  \href{http://dx.doi.org/10.1088/0264-9381/29/12/124016}{\emph{Class. Quant.
  Grav.} {\bf 29} (2012) 124016}, [\href{https://arxiv.org/abs/1202.0839}{{\tt
  1202.0839}}].

\bibitem{Audley:2017drz}
H.~Audley et~al., \emph{{Laser Interferometer Space Antenna}},
  \href{https://arxiv.org/abs/1702.00786}{{\tt 1702.00786}}.

\bibitem{Guo:2018npi}
Z.-K. Guo, R.-G. Cai and Y.-Z. Zhang, \emph{{Taiji Program: Gravitational-Wave
  Sources}},  \href{https://arxiv.org/abs/1807.09495}{{\tt 1807.09495}}.

\bibitem{Luo:2015ght}
{\scshape TianQin} collaboration, J.~Luo et~al., \emph{{TianQin: a space-borne
  gravitational wave detector}},
  \href{http://dx.doi.org/10.1088/0264-9381/33/3/035010}{\emph{Class. Quant.
  Grav.} {\bf 33} (2016) 035010}, [\href{https://arxiv.org/abs/1512.02076}{{\tt
  1512.02076}}].

\bibitem{Hobbs:2009yy}
G.~Hobbs et~al., \emph{{The international pulsar timing array project: using
  pulsars as a gravitational wave detector}},
  \href{http://dx.doi.org/10.1088/0264-9381/27/8/084013}{\emph{Class. Quant.
  Grav.} {\bf 27} (2010) 084013}, [\href{https://arxiv.org/abs/0911.5206}{{\tt
  0911.5206}}].

\bibitem{Carilli:2004nx}
C.~L. Carilli and S.~Rawlings, \emph{{Science with the Square Kilometer Array:
  Motivation, key science projects, standards and assumptions}},
  \href{http://dx.doi.org/10.1016/j.newar.2004.09.001}{\emph{New Astron. Rev.}
  {\bf 48} (2004) 979}, [\href{https://arxiv.org/abs/astro-ph/0409274}{{\tt
  astro-ph/0409274}}].

\bibitem{Hawking:1971ei}
S.~Hawking, \emph{{Gravitationally collapsed objects of very low mass}},
  {\emph{Mon. Not. Roy. Astron. Soc.} {\bf 152} (1971) 75}.

\bibitem{Carr:1974nx}
B.~J. Carr and S.~W. Hawking, \emph{{Black holes in the early Universe}},
  {\emph{Mon. Not. Roy. Astron. Soc.} {\bf 168} (1974) 399--415}.

\bibitem{Carr:1975qj}
B.~J. Carr, \emph{{The Primordial black hole mass spectrum}},
  \href{http://dx.doi.org/10.1086/153853}{\emph{Astrophys. J.} {\bf 201} (1975)
  1--19}.

\bibitem{Carr:2009jm}
B.~J. Carr, K.~Kohri, Y.~Sendouda and J.~Yokoyama, \emph{{New cosmological
  constraints on primordial black holes}},
  \href{http://dx.doi.org/10.1103/PhysRevD.81.104019}{\emph{Phys. Rev.} {\bf
  D81} (2010) 104019}, [\href{https://arxiv.org/abs/0912.5297}{{\tt
  0912.5297}}].

\bibitem{Carr:2016drx}
B.~Carr, F.~Kuhnel and M.~Sandstad, \emph{{Primordial Black Holes as Dark
  Matter}}, \href{http://dx.doi.org/10.1103/PhysRevD.94.083504}{\emph{Phys.
  Rev.} {\bf D94} (2016) 083504}, [\href{https://arxiv.org/abs/1607.06077}{{\tt
  1607.06077}}].

\bibitem{Carr:2017jsz}
B.~Carr, M.~Raidal, T.~Tenkanen, V.~Vaskonen and H.~Veermäe, \emph{{Primordial
  black hole constraints for extended mass functions}},
  \href{http://dx.doi.org/10.1103/PhysRevD.96.023514}{\emph{Phys. Rev.} {\bf
  D96} (2017) 023514}, [\href{https://arxiv.org/abs/1705.05567}{{\tt
  1705.05567}}].

\bibitem{Kashlinsky:2016sdv}
A.~Kashlinsky, \emph{{LIGO gravitational wave detection, primordial black holes
  and the near-IR cosmic infrared background anisotropies}},
  \href{http://dx.doi.org/10.3847/2041-8205/823/2/L25}{\emph{Astrophys. J.}
  {\bf 823} (2016) L25}, [\href{https://arxiv.org/abs/1605.04023}{{\tt
  1605.04023}}].

\bibitem{Bird:2016dcv}
S.~Bird, I.~Cholis, J.~B. Muñoz, Y.~Ali-Haïmoud, M.~Kamionkowski, E.~D.
  Kovetz et~al., \emph{{Did LIGO detect dark matter?}},
  \href{http://dx.doi.org/10.1103/PhysRevLett.116.201301}{\emph{Phys. Rev.
  Lett.} {\bf 116} (2016) 201301},
  [\href{https://arxiv.org/abs/1603.00464}{{\tt 1603.00464}}].

\bibitem{Clesse:2016vqa}
S.~Clesse and J.~García-Bellido, \emph{{The clustering of massive Primordial
  Black Holes as Dark Matter: measuring their mass distribution with Advanced
  LIGO}}, \href{http://dx.doi.org/10.1016/j.dark.2016.10.002}{\emph{Phys. Dark
  Univ.} {\bf 15} (2017) 142--147},
  [\href{https://arxiv.org/abs/1603.05234}{{\tt 1603.05234}}].

\bibitem{Sasaki:2016jop}
M.~Sasaki, T.~Suyama, T.~Tanaka and S.~Yokoyama, \emph{{Primordial Black Hole
  Scenario for the Gravitational-Wave Event GW150914}},
  \href{http://dx.doi.org/10.1103/PhysRevLett.117.061101}{\emph{Phys. Rev.
  Lett.} {\bf 117} (2016) 061101},
  [\href{https://arxiv.org/abs/1603.08338}{{\tt 1603.08338}}].

\bibitem{Gao:2018pvq}
T.-J. Gao and Z.-K. Guo, \emph{{Primordial Black Hole Production in
  Inflationary Models of Supergravity with a Single Chiral Superfield}},
  \href{https://arxiv.org/abs/1806.09320}{{\tt 1806.09320}}.

\bibitem{Cheng:2018yyr}
S.-L. Cheng, W.~Lee and K.-W. Ng, \emph{{Primordial black holes and associated
  gravitational waves in axion monodromy inflation}},
  \href{http://dx.doi.org/10.1088/1475-7516/2018/07/001}{\emph{JCAP} {\bf 1807}
  (2018) 001}, [\href{https://arxiv.org/abs/1801.09050}{{\tt 1801.09050}}].

\bibitem{Weinberg:2003ur}
S.~Weinberg, \emph{{Damping of tensor modes in cosmology}},
  \href{http://dx.doi.org/10.1103/PhysRevD.69.023503}{\emph{Phys. Rev.} {\bf
  D69} (2004) 023503}, [\href{https://arxiv.org/abs/astro-ph/0306304}{{\tt
  astro-ph/0306304}}].

\bibitem{Watanabe:2006qe}
Y.~Watanabe and E.~Komatsu, \emph{{Improved Calculation of the Primordial
  Gravitational Wave Spectrum in the Standard Model}},
  \href{http://dx.doi.org/10.1103/PhysRevD.73.123515}{\emph{Phys. Rev.} {\bf
  D73} (2006) 123515}, [\href{https://arxiv.org/abs/astro-ph/0604176}{{\tt
  astro-ph/0604176}}].

\bibitem{Saito}
R.~Saito and J.~Yokoyama, \emph{{Gravitational-Wave Constraints on the
  Abundance of Primordial Black Holes}},
  \href{http://dx.doi.org/10.1143/PTP.123.867}{\emph{Progress of Theoretical
  Physics} {\bf 123} (2010) 867--886},
  [\href{https://arxiv.org/abs/astro-ph.CO/0912.5317}{{\tt
  astro-ph.CO/0912.5317}}].

\bibitem{Seery.JoCaAP.2007.jan}
D.~Seery, J.~E. Lidsey and M.~S. Sloth, \emph{The inflationary trispectrum},
  \href{http://dx.doi.org/10.1088/1475-7516/2007/01/027}{\emph{Journal of
  Cosmology and Astroparticle Physics} {\bf 2007} (2007) 027--027},
  [\href{https://arxiv.org/abs/astro-ph/0610210v2}{{\tt astro-ph/0610210v2}}].

\bibitem{Byrnes.PRD.2006.11}
C.~T. Byrnes, M.~Sasaki and D.~Wands, \emph{Primordial trispectrum from
  inflation}, \href{http://dx.doi.org/10.1103/PhysRevD.74.123519}{\emph{Phys.
  Rev. D} {\bf 74} (2006) 123519},
  [\href{https://arxiv.org/abs/astro-ph/0611075v2}{{\tt astro-ph/0611075v2}}].

\bibitem{Barnacka:2012bm}
A.~Barnacka, J.~F. Glicenstein and R.~Moderski, \emph{{New constraints on
  primordial black holes abundance from femtolensing of gamma-ray bursts}},
  \href{http://dx.doi.org/10.1103/PhysRevD.86.043001}{\emph{Phys. Rev.} {\bf
  D86} (2012) 043001}, [\href{https://arxiv.org/abs/1204.2056}{{\tt
  1204.2056}}].

\bibitem{Katz:2018zrn}
A.~Katz, J.~Kopp, S.~Sibiryakov and W.~Xue, \emph{{Femtolensing by Dark Matter
  Revisited}},
  \href{http://dx.doi.org/10.1088/1475-7516/2018/12/005}{\emph{JCAP} {\bf 1812}
  (2018) 005}, [\href{https://arxiv.org/abs/1807.11495}{{\tt 1807.11495}}].

\bibitem{Graham:2015apa}
P.~W. Graham, S.~Rajendran and J.~Varela, \emph{{Dark Matter Triggers of
  Supernovae}}, \href{http://dx.doi.org/10.1103/PhysRevD.92.063007}{\emph{Phys.
  Rev.} {\bf D92} (2015) 063007}, [\href{https://arxiv.org/abs/1505.04444}{{\tt
  1505.04444}}].

\bibitem{Niikura:2017zjd}
H.~Niikura, M.~Takada, N.~Yasuda, R.~H. Lupton, T.~Sumi, S.~More et~al.,
  \emph{{Microlensing constraints on primordial black holes with the Subaru/HSC
  Andromeda observation}},  \href{https://arxiv.org/abs/1701.02151}{{\tt
  1701.02151}}.

\bibitem{Allsman:2000kg}
{\scshape Macho} collaboration, R.~A. Allsman et~al., \emph{{MACHO project
  limits on black hole dark matter in the 1-30 solar mass range}},
  \href{http://dx.doi.org/10.1086/319636}{\emph{Astrophys. J.} {\bf 550} (2001)
  L169}, [\href{https://arxiv.org/abs/astro-ph/0011506}{{\tt
  astro-ph/0011506}}].

\bibitem{Tisserand:2006zx}
{\scshape EROS-2} collaboration, P.~Tisserand et~al., \emph{{Limits on the
  Macho Content of the Galactic Halo from the EROS-2 Survey of the Magellanic
  Clouds}}, \href{http://dx.doi.org/10.1051/0004-6361:20066017}{\emph{Astron.
  Astrophys.} {\bf 469} (2007) 387--404},
  [\href{https://arxiv.org/abs/astro-ph/0607207}{{\tt astro-ph/0607207}}].

\bibitem{Wyrzykowski:2011tr}
L.~Wyrzykowski et~al., \emph{{The OGLE View of Microlensing towards the
  Magellanic Clouds. IV. OGLE-III SMC Data and Final Conclusions on MACHOs}},
  \href{http://dx.doi.org/10.1111/j.1365-2966.2011.19243.x}{\emph{Mon. Not.
  Roy. Astron. Soc.} {\bf 416} (2011) 2949},
  [\href{https://arxiv.org/abs/1106.2925}{{\tt 1106.2925}}].

\bibitem{Brandt:2016aco}
T.~D. Brandt, \emph{{Constraints on MACHO Dark Matter from Compact Stellar
  Systems in Ultra-Faint Dwarf Galaxies}},
  \href{http://dx.doi.org/10.3847/2041-8205/824/2/L31}{\emph{Astrophys. J.}
  {\bf 824} (2016) L31}, [\href{https://arxiv.org/abs/1605.03665}{{\tt
  1605.03665}}].

\bibitem{Ali-Haimoud:2016mbv}
Y.~Ali-Haïmoud and M.~Kamionkowski, \emph{{Cosmic microwave background limits
  on accreting primordial black holes}},
  \href{http://dx.doi.org/10.1103/PhysRevD.95.043534}{\emph{Phys. Rev.} {\bf
  D95} (2017) 043534}, [\href{https://arxiv.org/abs/1612.05644}{{\tt
  1612.05644}}].

\bibitem{Cai:2018rqf}
R.-G. Cai, T.-B. Liu and S.-J. Wang, \emph{{Sensitivity of primordial black
  hole abundance on the reheating phase}},
  \href{http://dx.doi.org/10.1103/PhysRevD.98.043538}{\emph{Phys. Rev.} {\bf
  D98} (2018) 043538}, [\href{https://arxiv.org/abs/1806.05390}{{\tt
  1806.05390}}].

\bibitem{Wang:2019kaf}
S.~Wang, T.~Terada and K.~Kohri, \emph{{Prospective constraints on the
  primordial black hole abundance from the stochastic gravitational-wave
  backgrounds produced by coalescing events and curvature perturbations}},
  \href{https://arxiv.org/abs/1903.05924}{{\tt 1903.05924}}.

\bibitem{Harada:2013epa}
T.~Harada, C.-M. Yoo and K.~Kohri, \emph{{Threshold of primordial black hole
  formation}}, \href{http://dx.doi.org/10.1103/PhysRevD.88.084051,
  10.1103/PhysRevD.89.029903}{\emph{Phys. Rev.} {\bf D88} (2013) 084051},
  [\href{https://arxiv.org/abs/1309.4201}{{\tt 1309.4201}}].

\end{thebibliography}\endgroup

\end{document}